\documentclass[11pt]{article}
\usepackage[margin=1in]{geometry}
\usepackage{rpmacros}
\usepackage[T1]{fontenc}
\usepackage[usenames,dvipsnames]{color}
\usepackage{stmaryrd}
\usepackage{lineno}
\usepackage{xfrac}
\usepackage{mathpazo}
\usepackage{todonotes}
\usepackage{setspace}
\usepackage{nicefrac}
\usepackage{gitinfo}
\usepackage{float}
\usepackage{scrextend}
\allowdisplaybreaks

\usepackage{fullpage}
\usepackage[utf8]{inputenc}
\usepackage[russian,english]{babel}

\usepackage{color}
\usepackage[
pagebackref, 
pdfstartview=FitH,pdfpagemode=UseNone,colorlinks=true,citecolor=blue,linkcolor=blue]{hyperref}
\hypersetup{
  linkcolor=[rgb]{0.3,0.3,0.6},
  citecolor=[rgb]{0.2, 0.6, 0.2},
  urlcolor=[rgb]{0.6, 0.2, 0.2}
}
\usepackage[nameinlink,capitalise]{cleveref}

\usepackage{amsthm}
\usepackage{thmtools,thm-restate}

\numberwithin{equation}{section}
\declaretheoremstyle[bodyfont=\it,qed=\qedsymbol]{noproofstyle}

\declaretheorem[name=Observation,numbered=no]{observation*}

\declaretheorem[numberlike=equation]{theorem}

\declaretheorem[name=Theorem,numbered=no]{theorem*}

\declaretheorem[numberlike=equation]{lemma}
\declaretheorem[name=Lemma,numbered=no]{lemma*}

\declaretheorem[name=Corollary,numbered=no]{corollary*}

\declaretheorem[numberlike=equation]{proposition}
\declaretheorem[name=Proposition,numbered=no]{proposition*}

\declaretheorem[numberlike=equation]{claim}
\declaretheorem[name=Claim,numbered=no]{claim*}

\declaretheorem[name=Conjecture,numbered=no]{conjecture*}

\declaretheorem[name=Question,numbered=no]{question*}

\declaretheoremstyle[bodyfont=\it,qed=$\lrcorner$]{defstyle} 

\declaretheorem[numberlike=equation,style=defstyle]{definition}
\declaretheorem[unnumbered,name=Definition,style=defstyle]{definition*}

\declaretheorem[unnumbered,name=Example,style=defstyle]{example*}

\declaretheorem[unnumbered,name=Notation=defstyle]{notation*}

\declaretheorem[unnumbered,name=Construction,style=defstyle]{construction*}

\declaretheorem[numberlike=equation,style=defstyle]{remark}
\declaretheorem[unnumbered,name=Remark,style=defstyle]{remark*}

\usepackage{nth}
\usepackage{intcalc}
\usepackage{etoolbox}
\usepackage{xstring}

\usepackage{ifpdf}
\ifpdf
\else
\usepackage[quadpoints=false]{hypdvips}
\fi

\newcommand{\ECCCurl}[2]{http://eccc.hpi-web.de/report/\ifnumcomp{#1}{>}{93}{19}{20}#1/#2/}

\newcommand{\shortECCC}[2]{\texttt{\href{http://eccc.hpi-web.de/report/\ifnumcomp{#1}{>}{93}{19}{20}#1/#2/}{eccc:TR#1-#2}}}

\newcommand{\parseECCC}[1]{
\StrSubstitute{#1}{TR}{}[\tmpstring]%
\IfSubStr{\tmpstring}{/}{ 
\StrBefore{\tmpstring}{/}[\ecccyear]%
\StrBehind{\tmpstring}{/}[\ecccreport]%
}{
\StrBefore{\tmpstring}{-}[\ecccyear]%
\StrBehind{\tmpstring}{-}[\ecccreport]%
}%
\shortECCC{\ecccyear}{\ecccreport}}

\usetikzlibrary{decorations.pathreplacing,arrows}

\newif\ifnote
\notetrue
\ifnote
\newcommand{\PHnote}[1]{\textcolor{BrickRed}{\guillemotleft PH: #1\guillemotright}}
\newcommand{\MKnote}[1]{\textcolor{Purple}{\guillemotleft MK: #1\guillemotright}}
\newcommand{\MSnote}[1]{\textcolor{OliveGreen}{\guillemotleft MS: #1\guillemotright}}
\newcommand{\SBnote}[1]{\textcolor{NavyBlue}{\guillemotleft SB: #1\guillemotright}}
\else
\newcommand{\PHnote}[1]{}
\newcommand{\MKnote}[1]{}
\newcommand{\MSnote}[1]{}
\newcommand{\SBnote}[1]{}
\fi

\newtheorem*{SZquestion}{Algorithmic SZ question}
\newtheorem*{mainthm}{Main result}

\newcommand{\hpartial}{{\mathchar'26\mkern-10mu \partial}}

\newcommand{\ehref}[1]{\href{mailto:#1}{#1}}

\newcommand{\D}{\Delta}

\newcommand{\enc}{{\mathsf{Enc}}}
\newcommand{\eval}{{\mathsf{Eval}}}

\let\epsilon\varepsilon
\newcommand{\coeff}{\operatorname{coeff}}
\newcommand{\mult}{\operatorname{mult}}
\newcommand{\dist}{\operatorname{Dist}}

\newcommand{\vy}{\vecy}
\newcommand{\vz}{\vecz}
\newcommand{\vu}{\vecu}
\newcommand{\vf}{\vecf}
\newcommand{\vx}{\vecx}
\newcommand{\ve}{\vece}

\newcommand{\va}{\veca}
\newcommand{\vb}{\vecb}
\newcommand{\vc}{\vecc}

\onehalfspace

\title{Decoding Multivariate Multiplicity Codes on Product Sets}

\author{ 
{Siddharth Bhandari\thanks{Tata Institute of Fundamental Research, Mumbai, India. \ehref{\{siddharth.bhandari,prahladh\}@tifr.res.in}. Research supported by the Department of Atomic Energy, Government of India, under project no. 12-R\&D-TFR-5.01-0500 and in part by the Google PhD Fellowship and Swarnajayanti fellowship.}}
\and 
{Prahladh Harsha\samethanks}
\and
{Mrinal Kumar\thanks{ Department of Computer Science \& Engineering,
    IIT Bombay, Mumbai, India. \ehref{mrinal@cse.iitb.ac.in}. }}
\and
{Madhu Sudan\thanks{School of Engineering and Applied Sciences,
    Harvard University, Cambridge, MA, USA. \ehref{madhu@cs.harvard.edu}. Supported in part by a Simons Investigator Award and NSF Award CCF 1715187.}}
}

\date{}

\begin{document}

\maketitle

\begin{abstract}
The multiplicity Schwartz-Zippel lemma bounds the total multiplicity of zeroes of a multivariate polynomial on a product set. This lemma motivates the multiplicity codes of Kopparty, Saraf and Yekhanin [J. ACM, 2014], who showed how to use this lemma to construct high-rate locally-decodable codes. However, the algorithmic results about these codes crucially rely on the fact that the polynomials are evaluated on a vector space and not an arbitrary product set. 

In this work, we show how to decode multivariate multiplicity codes of large multiplicities in polynomial time over finite product sets (over fields of large characteristic and zero characteristic).  Previously such decoding algorithms were not known even for a positive fraction of errors. In contrast, our work goes all the way to the distance of the code and in particular exceeds both the unique decoding bound and the Johnson bound. For errors exceeding the Johnson bound, even combinatorial list-decodablity of these codes was not known.

Our algorithm is an application of the classical polynomial method directly to the multivariate setting. In particular, we do not rely on a reduction from the multivariate to the univariate case as is typical of many of the existing results on decoding codes based on multivariate polynomials.  However, a vanilla application of the polynomial method in the multivariate setting does not yield a polynomial upper bound on the list size. We obtain a polynomial bound on the list size by taking an alternative view of multivariate multiplicity codes. In this view,  we glue all the partial derivatives of the same order  together  using a fresh set $\vz$ of variables.  We then apply the polynomial method by viewing this as a problem over the field $\F(\vz)$ of rational functions in $\vz$. 

\end{abstract}

\newpage

\section{Introduction}
The classical Schwartz-Zippel Lemma (due to Ore~\cite{Ore1922}, Schwartz~\cite{Schwartz1980}, Zippel~\cite{Zippel1979} and DeMillo \& Lipton~\cite{DemilloL1978}) states that if $\F$ is a field, and $f \in \F[x_1, x_2, \ldots, x_k]$ is a \emph{non-zero} polynomial of degree $d$, and $S\subseteq \F$ is an arbitrary finite subset of $\F$, then the number of points on the grid \footnote{We use ``grids'' and ``product sets'' interchangeably (see also \cref{rem:grid}).} $S^k$  where $f$ is zero is upper bounded by $d|S|^{k-1}$. A higher order {\em multiplicity} version of this lemma (due to Dvir, Kopparty, Saraf and Sudan~\cite{DvirKSS2013}) states the number of points on the grid $S^k$ where $f$ is zero with \emph{multiplicity}\footnote{This means that all the partial derivatives of $f$ of order at most $s-1$ are zero at this point. See \cref{sec: prelims} for a formal definition.} at least $s$ is upper bounded by $\frac{d|S|^{k-1}}{s}$.\footnote{This bound is only interesting when $|S| > d/s$ so that $\frac{d|S|^{k-1}}{s}$ is less than the trivial bound of $|S|^{k}$.} 

This innately basic statement about low degree polynomials has had innumerable applications in both theoretical computer science and discrete mathematics and has by now become a part of the standard toolkit when working with low degree polynomials~\cite{Saraf2011,Guth2016}. Despite this, the following natural algorithmic version of this problem remains open. 

\begin{SZquestion}\label{q: starting question}
Let $\F$ be a field, and $S$, $d$, $k$ be as above. Design an efficient algorithm that takes as input an arbitrary function $P: S^{k} \to \F^{\binom{s + k -1}{k}}$ and finds a polynomial $f \in \F[x_1, x_2, \ldots, x_k]$ of degree at most $d$ (if one exists) such that the function $\enc(f): S^{k} \to \F^{\binom{s + k-1}{k}}$ defined as 
\[
\enc(f)(\va) = \left(\frac{\partial f}{\partial \vx^{\ve}}(\va) : \deg(\vx^{\ve}) < s \right) 
\]
differs from $P$ on less than $\frac12\left(1-\frac{d}{s|S|}\right)$ fraction of points on $S^k$.   
\end{SZquestion}
The aforementioned multiplicity Schwartz-Zippel lemma (henceforth, referred to as the multiplicity SZ lemma for brevity)
assures us that if there is a polynomial $f \in \F[x_1, x_2, \ldots, x_k]$ such that $\enc{(f)}$ differs from $P$ on less than  $\frac12\left(1-\frac{d}{s|S|}\right)$ fraction of points, then it must be unique! Thus, in some sense, the above question is essentially asking for an algorithmic version of the multiplicity SZ lemma. 

Although a seemingly natural problem, especially given the ubiquitous presence of the SZ lemma in computer science, this question continues to remain open for even bivariate polynomials! In fact, even the $s = 1$ case, which corresponds to an algorithmic version of the classical SZ lemma (without multiplicities) was only very recently resolved in a beautiful work of Kim and Kopparty~\cite{KimK2017}. Unfortunately, their algorithm does not seem to extend to the case of $s > 1$, and they mention this as one of the open problems. 

In this work, we make some progress towards answering the algorithmic SZ question. In particular, we design an efficient deterministic algorithm for this problem when the field $\F$ has characteristic zero or larger than the degree $d$, the dimension $k$ is an arbitrary constant and the multiplicity parameter $s$ is a sufficiently large constant. In fact, in this setting we prove a  stronger result, which we now informally state (see \cref{thm:main} for a formal statement). 



\begin{mainthm}\label{thm: intro}
  Let $\epsilon \in (0,1)$ be an arbitrary constant, $k\in \N$ be a positive constant and $s$ be a large enough positive integer.
  Over fields $\F$ of characteristic zero or characteristic larger than $d$, there is a deterministic polynomial algorithm that on input $P$ outputs all degree $d$ polynomials $f \in \F[x_1, x_2, \ldots, x_k]$ such that $\enc(f)$ differs from the input function $P: S^{k} \to \F^{\binom{s + k -1}{k}}$ on  less than  $\left(1-\frac{d}{s|S|} -\epsilon\right) $ fraction of points on the grid $S^k$. 
\end{mainthm}
We note that the fraction of errors that can be tolerated in the above result is $1-\frac{d}{s|S|}-\epsilon$, which is significantly larger than the error parameter in the algorithmic SZ question. Therefore, we no longer have the guarantee of a unique solution $f$ such that the function $\enc(f)$ which is \emph{close} to $P$. In fact, for this error regime, it is not even clear that the number of candidate solutions is polynomially bounded. The algorithm stated in the main result outputs all such candidate solutions, and in particular, shows that their number is polynomially bounded (for constant $k$). This fraction of errors is the best one can hope for since there are functions $P$ (for instance, the all zero's function) which have super-polynomially many polynomials of degree $d$ which are $\left(1 - \frac{d}{s|S|}\right) $-close to $P$. (see \cref{sec: super polynomial min weight}).  

In the language of error correcting codes, the algorithmic SZ question is the question of designing efficient unique decoding algorithms for multivariate multiplicity codes over arbitrary product sets when the error is at most half the minimum distance, and main result gives an efficient algorithm for the possibly harder problem of list decoding these codes from relative error $\delta - \epsilon$, where $\delta := 1-\frac{d}{s|S|}$ is the distance of the code, provided that the field has characteristic larger than $d$ or zero, $k$ is a constant  and $s$ is large enough. In the next section, we define some of these notions, state and discuss the results and the prior work in this language.


\subsection{Multiplicity codes}
Polynomial based error correcting codes, such as the Reed-Solomon codes and Reed-Muller codes, are a very important family of codes in coding theory both in theory and practice. Multiplicity codes are a natural generalization of Reed-Muller codes wherein at each evaluation point, one not only gives the evaluation of the polynomial $f$, but also all its derivatives up to a certain order. 


Formally, let $\F$ be a field, $s$ a positive integer, $S \subset \F$ an arbitrary subset of the field $\F$, $d \leq s|S|$ the degree parameter and $k \geq 1$ the ambient dimension. 
The codewords of the $k$-variate order-$s$ multiplicity code of degree-$d$ polynomials over $\F$ on the grid $S^k$ is obtained by evaluating a $k$-variate polynomial of total degree at most $d$, along with all its derivatives of order less than $s$ at all points in the grid $S^k$. Thus, a codeword corresponding to the polynomial $f$ of total degree at most $k$ can be viewed as a function $\enc_{s, S}(f) \colon S^k \to \F^{|E|}$ where $E := \{ \ve \in \Z_{\geq 0}^k \mid 0 \leq \|\ve \|_1 < s\}$ and
\[
  \enc_{s, S}(f)|_{\va} = \left(\frac{\hpartial f}{\hpartial \vx^\ve}(\va) : \ve \in E\right)
   \]
 where $\frac{\hpartial f}{\hpartial \vx^\ve}$is the Hasse derivative of the polynomial $f$ with respect to $\vx^\ve$.
The $s=1$ version of these multiplicity codes corresponds to the classical Reed-Solomon codes (univariate case, $k=1$) and Reed-Muller codes (multivariate setting, $k >1$). The distance of these codes is $\delta := 1- \frac{d}{s|S|}$, which follows from the multiplicity SZ Lemma mentioned earlier in the introduction.

Univariate multiplicity codes were first studied by Rosenbloom \& Tsfasman \cite{RosenbloomT1997} and Nielsen \cite{Nielsen2001}. Multiplicity codes for general $k$ and $s$ were introduced by Kopparty, Saraf and Yekhanin \cite{KoppartySY2014} in the context of local decoding. Subsequently, Kopparty~\cite{Kopparty2015} and Guruswami \& Wang~\cite{GuruswamiW2013} independently proved that the univariate multiplicity codes over prime fields (or more generally over fields whose characteristic is larger than the degree of the underlying polynomials) achieve ``list-decoding capacity''. In the same work, Kopparty~\cite{Kopparty2015} proved that multivariate multiplicity codes were list decodable up to the Johnson bound.



We remark that in the case of univariate multiplicity codes (both Reed-Solomon and larger order multiplicity codes), the decoding algorithms work for all choices of the set $S \subset \F$. However, all decoding algorithms for the multivariate setting (both Reed-Muller and larger order multiplicity codes) work only when the underlying set $S$ has a nice algebraic structure (eg., $S = \F$) or when the degree $d$ is very small (cf, the Reed-Muller list-decoding algorithm of Sudan~\cite{Sudan1997} and its multiplicity variant due to Guruswami \& Sudan \cite{GuruswamiS1999}). The only exception to this is the unique decoding algorithm of Kim and Kopparty~\cite{KimK2017} of Reed-Muller codes over product sets.

\subsection{Our results}

Below we state and contrast our results on the problem of decoding multivariate multiplicity codes (over grids) from a $\delta-\epsilon$ fraction of errors for any constant $\epsilon \in (0,1)$ where $\delta$ is the distance of the code. Our first result is as follows.

\begin{theorem}[List decoding of multivariate multiplicity codes with polynomial list size]\label{thm:main}
For every $\epsilon \in (0,1)$ and integer $k$, there exists an
integer $s_0$ such that for all $s \geq s_0$, degree parameter $d$, fields $\F$ of size $q$
and characteristic larger than $d$, and any set $S \subseteq \F$ where $d <
s|S|$, the following holds. 

For $k$-variate order-$s$ multiplicity code of degree-$d$ polynomials over $\F$ on the grid $S^k$, there is an efficient algorithm which when given a received word $P$ , outputs all code words with agreement at least $(1-\delta + \epsilon)$ with $P$, where $\delta = 1 - d/(s|S|)$ is the relative distance of this code.
\end{theorem}

\begin{remark}\label{rem:grid}
A general product set in $\F^k$ is of the form $S_1 \times S_2 \times \cdots S_k$, where each $S_i$ is a subset of $\F$. For the ease of notation, we always work with product sets which are grids $S^k$ for some $S \subseteq \F$ even though all of our results hold for general product sets. 
\end{remark}

As indicated before, this is the best one can hope for with respect to polynomial time list-decoding algorithms for multiplicity codes since there are super-polynomially many codewords with minimum distance $\delta = 1 - d/(s|S|)$ (see \cref{sec: super polynomial min weight}). Till recently, it was not known if multivariate multiplicity codes were list decodable beyond the Johnson bound (even for the case $S = \F$). 
For the case of grids $S^k$, where $S \subseteq \F$ is an arbitrary set, even unique decoding algorithms were not known. We note that the above result does not yield a list-decoding algorithm for all multiplicities, but only for large enough multiplicities (based on the dimension $k$ and the error parameter $\epsilon$).

Kopparty, Ron-Zewi, Saraf and Wootters~\cite{KoppartyRSW2018} showed how to reduce the size of the list for univariate multiplicity codes from polynomial to constant (dependent only on the error parameter $\epsilon$). We use similar ideas, albeit in the multivariate setting, to reduce the list size in \cref{thm:main} to  constant (dependent only on the error parameter $\epsilon$ and the dimension $k$).

\begin{theorem}[List decoding of multivariate multiplicity codes  with constant list size]\label{thm: constant list size}
For every $\epsilon \in (0,1)$ and integer $k$, there exists an
integer $s_0$ such that for all $s \geq s_0$, degree parameter $d$, fields $\F$ of size $q$
and characteristic larger than $d$, and any set $S \subseteq \F$ where $d <
s|S|$, the following holds. 

For $k$-variate order-$s$ multiplicity code of degree-$d$ polynomials over $\F$ on the grid $S^k$, there is a randomized algorithm which requires $\poly\left(d^{k^2}, |S|^{k^2}, \exp\left({O\left(\frac{k^2}{\epsilon}\log^3 \frac{1}{\epsilon}\right)}\right)\right)$ operations over the field $\F$ and 
which when given a received word $P$, outputs all code words with
agreement at least $(1-\delta + \epsilon)$ with $P$, where $\delta =
1 - d/(s|S|)$ is the relative distance of this code. 

Moreover, the number of such codewords is at most $ \exp\left({O\left(\frac{k^2}{\epsilon}\log^2 \frac{1}{\epsilon}\right)}\right)$.
\end{theorem}

\begin{remark}\label{rmk: improved running time intro}
We remark that by taking a slightly different view of the list decoding algorithm \autoref{thm:main} and \autoref{thm: constant list size}, the upper bound on the number of field operations needed  in \autoref{thm:main} and \autoref{thm: constant list size} can be improved to $\poly(|S|^k, d^k)$. We sketch this view in \autoref{sec: alternative view of the algorithm} and note the runtime analysis in \autoref{rmk: improved running time}.
\end{remark}

The above two results are a generalization (and imply) the corresponding theorems for the univariate setting due to Kopparty~\cite{Kopparty2015} and Guruswami \& Wang~\cite{GuruswamiW2013} and Kopparty, Ron-Zewi, Saraf \& Wootters~\cite{KoppartyRSW2018}. We remark that  Kopparty, Ron-Zewi , Saraf and Wootters~\cite{KoppartyRSW2018} in the recent improvement to their earlier work prove a similar list-decoding algorithm for multivariate multiplicity codes as \cref{thm: constant list size} for the case when $S = \F$. Though their list-decoding algorithm does not extend to products sets, it has the added advantage that it is {\em local}. 

As noted earlier the only previous algorithmic method for decoding
polynomial-based codes over product sets was that of Kim and
Kopparty~\cite{KimK2017}. We describe the ideas in our algorithm shortly (in \cref{sec: proof overview}), but stress here that our approach is very different from that of Kim and Kopparty. Their
work may be viewed as an algorithmic version of the inductive proof of
the SZ lemma, and indeed recovers the SZ lemma as a consequence. Their
work uses algorithmic aspects of algebraic decoding as a black box (to
solve univariate cases).  Our work, in contrast, only relies on the multiplicity SZ lemma as a black box. Instead, we  open up the "algebraic decoding''
black box and make significant changes there, thus adding to the toolkit
available to deal with polynomial evaluations over product sets.



\subsection{Further discussion and open problems}

Our result falls short of completely resolving the algorithmic SZ question in two respects; though it works for all dimensions $k$ it only works when the multiplicity parameter $s$ is large enough and when the characteristic of the field is either zero or larger than the degree parameter. Making improvements on any of these fronts is an interesting open problem.
\begin{description}
\item[All multiplicities:] The algorithms presented in this paper decode all the way up to distance if the multiplicity parameter $s$ is large enough. However, for small multiplicities, even the unique decoding problem is open. For $s=1$, the result due to Kim and Kopparty~\cite{KimK2017} addresses the unique decoding question, but the list-decoding question for product sets is open.
\item[Fields of small characteristic:] All known proofs of list-decoding multiplicity codes beyond the Johnson bound (both algorithmic and combinatorial) require the field to be of zero characteristic or large enough characteristic. The problem of list-decoding multiplicity codes over small characteristic beyond the Johnson bound is open even for the univariate setting. As pointed to us by Swastik Kopparty, this problem of list-decoding univariate multiplicity codes over fields of small characteristic beyond the Johnson bound is intimately related to list-decoding Reed-Solomon codes beyond the Johnson bound.
\end{description}

For a more detailed discussion of multiplicity codes and related open problems, we refer the reader to the excellent survey by Kopparty~\cite{Kopparty2014}. 

\subsection*{Organization} The rest of this paper is organized as follows. We begin with an overview of our proofs in \cref{sec: proof overview} followed by some preliminaries (involving Hasse derivatives, their properties, multiplicity codes) in \cref{sec: prelims}. We then describe and analyze the list-decoding algorithm for multivariate multiplicity codes in \cref{sec:polylist}, thus proving \cref{thm:main}. In \cref{sec: constant list size}, we then show how to further reduce the list-size to a constant, thus proving \cref{thm: constant list size}. In \cref{sec: wronskian}, we prove some properties of subspace restriction of multivariate multiplicity codes needed in \cref{sec: constant list size}. In \cref{sec:  super polynomial min weight}, we show that there are super-polynomially many minimum-weight codewords, thus proving the tightness of \cref{thm:main,thm: constant list size} with respect to list-decoding radius.

\section{Proof overview}\label{sec: proof overview}

In this section, we first describe some of the hurdles in extending the univariate algorithms of Kopparty~\cite{Kopparty2015} and Guruswami \& Wang~\cite{GuruswamiW2013} to the multivariate setting, especially for product sets and then given a detailed overview of the proofs of \cref{thm:main} and \cref{thm: constant list size}.

\subsection{Background and motivation for our algorithm}

To explain our algorithm, it will be convenient to recall the general polynomial method framework underlying the list-decoding algorithms in the univariate setting  due to Kopparty~\cite{Kopparty2015} and Guruswami \& Wang~\cite{GuruswamiW2013}. . Let $P : S \to \F^{s}$ be the received word and $1 \leq m \leq s$

\begin{description}
\item[Step 1: Algebraic Explanation.] Find a polynomial $Q(x,y_1,\dots, y_m) \in \F[x,y_1,\dots, y_m]$ of appropriate degree constraints that ``explains'' the received word $P$.
\item[Step 2: $Q$ contains the close codewords.] Show that every low-degree polynomial $f$ whose encoding agrees with $P$ in more than $(1-\delta+\epsilon)$-fraction of points satisfies the following condition.
  \[ Q\left(x, f(x), \frac{\hpartial f}{\hpartial x}, \frac{\hpartial f}{\hpartial x^2}, \dots, \frac{\hpartial f}{\hpartial x^{m-1}}\right) = 0\, .\]
\item[Step 3: Reconstruction step.] Recover every polynomial $f$ that satisfies the above condition.
  \end{description}

  The main (and only) difference between the list-decoding algorithms of Kopparty~\cite{Kopparty2015} and Guruswami \& Wang~\cite{GuruswamiW2013} is that Guruswami and Wang show that it suffices to work with a polynomial $Q$ which is linear in the $y$-variables, more precisely, $Q(x, y_1,\dots, y_m)$ of the form $Q_0(x)+Q_1(x) \cdot y_1 +\dots + Q_m(x)\cdot y_m$, while Kopparty allows for larger degrees in the $y$-variables. As a result, Kopparty performs the recovery step by solving a differential equation while Guruswami and Wang observe that dueto  the simple structure of $Q$, the solution can be obtained by solving a linear system of equations.

  How is multivariate list-decoding performed? There are by now two standard approaches. Inspired by the Pellikaan-Wu~\cite{PellikaanW2004} observation that Reed-Muller codes are a subcode of Reed-Solomon codes over an extension field, Kopparty performs a similar reduction of the multivariate multiplicity code to a univariate multiplicity code over an extension field. Another approach is to solve the multivariate case by solving the univariate subproblem on various lines in the space.  
However, both these approaches  work only if the set $S = \F$ or has some special algebraic structure. 

  For our proof, we  take an alternate approach and always work in the multivariate setting without resorting to a reduction to the univariate setting. As we shall see, our approach has some advantages over that of Kopparty~\cite{Kopparty2015}, both in quantitative terms, since the algorithm can tolerate a larger number of errors, and in qualitative terms, since the underlying set of evaluation points does not have to be an algebraically nice subset of $\F^k$ as in \cite{Kopparty2015}; evaluations on an arbitrary grid suffice for the algorithm to work.

  To extend the univariate list-decoding algorithm outlined above to the multivariate setting, we adopt the following approach. We consider a new set of formal variables $\vz$ and instead of directly working with the information about partial derivatives in the received word, we think of the partial derivatives of the same order as being \emph{glued} together using monomials in $\vz$. With this reorganized (and somewhat mysterious) view of the partial derivatives, we follow the outline of the univariate setting as described above.  
We find a polynomial $Q$ with coefficients from the field of fractions $\F(\vz)$ instead of just $\F$ in the interpolation step to explain the received word $P$. Thus, in this instance, the linear system in the interpolation step is over the field $\F(\vz)$. We then argue that $Q$ \emph{contains} information about all the codewords that are close to the received word, and eventually \emph{solve} $Q$ to recover all the codewords close to the received word.  This might seem rather strange to begin with, but these ideas of gluing together the partial derivatives and working over the field $\F(\vz)$ immediately generalize the univariate list decoding algorithm to the multivariate setting. Working with this field of fractions $\F(\vz)$ comes with its costs; it makes some of the steps costly and in particular, the recovery step far more elaborate than that in the Guruswami-Wang setting. However, this recovery step happens to be a special case of similar step in the recent work of Guo, Kumar, Saptharishi and Solomon~\cite{GuoKSS2019} and we adapt their algorithm to our setting.

As a first attempt, a more standard way to generalize the algorithms of Kopparty \cite{Kopparty2015} and Guruswami \& Wang \cite{GuruswamiW2013} to the multivariate setting would have been  to work with the partial derivatives directly. And, while this approach seems alright for the interpolation step, it seems hard to work with when we try to solve the resulting equation to recover all the close enough codewords. In particular, it isn't even clear in this set up that the number of solutions of the algebraic explanation (and hence, the number of close enough codewords) is polynomially bounded. This mysterious step of gluing together derivatives of the same order in a reversible manner (in the sense that we can read off the individual derivatives from the glued term) gets around this problem, and makes it viable to prove a polynomial upper bound on the number of solutions, and eventually solve the equation to recover all the close enough codewords. 


Given this background, we now give a more detailed outline of our algorithm below. 

\subsection{\cref{thm:main} : Multivariate list-decoding algorithm with polynomial-sized lists}

\subsubsection*{Viewing the encoding as a formal power series}
Multiplicity codes are described by saying that the encoding of a polynomial $f \in \F[\vx]$ consists of the evaluation of \emph{all} partial derivatives of $f$ of order at most $s-1$ at every point in the appropriate evaluation set, e.g. the grid $S^k$. For our algorithm, we think of these partial derivatives of $f$ as being rearranged on the basis of the order of the derivatives as follows. We take a fresh set of formal variables $\vz$ and define the following differential operators. 
\[
\D_i(f) := \sum_{\ve: \|\ve\|_1= i} \vz^{\ve} \cdot \frac{\hpartial f(\vx)}{\hpartial \vx^{\ve}}  
\]
where  $\frac{\hpartial f}{\hpartial \vx^\ve}$ denotes the Hasse derivative\footnote{Since we have both $\vx$ and $\vz$ variables, we use the notation $\frac{\hpartial f}{\hpartial x}$ to denote the Hasse derivative wrt variable $x$ to explicitly indicate which variable the derivative is being taken} of the polynomial $f$ with respect to $\vx^\ve$.

Let $\D(f)$ be an $s$ tuple of polynomials defined as follows. 
\[
\D(f) := \left(\D_0(f), \D_1(f), \ldots, \D_{s-1}(f)\right) \, . 
\]
We view the encoding for $f$ as giving us the evaluation of the tuple $\D(f) \in \F[\vx, \vz]$ as $\vx$ varies in $S^k$. Note that for every fixing of $\vx$ to some $\va \in S^k$,  $\D(f)(\va)$ is in $\F[\vz]^s$. Thus, the alphabet size is still large. Clearly, this is just a change of viewpoint, as we can go from the original encoding to this and back efficiently, and at this point it is unclear that this change of perspective  would be useful. 

\subsubsection*{Finding an equation satisfied by all close enough codewords}
Let $P$ be a received word. We view $P$ as a function $P:S^k\to \Sigma^s$, where $\Sigma = \F[\vz]$, as discussed in the previous step. The goal of the decoding step is to find all the polynomials $f \in \F[\vx]$ of degree at most $d$, whose encoding is close enough to $P$.

As a first step towards this, we find a non-zero polynomial $Q(\vx, \vy) \in \F(\vz)[\vx, \vy]$ of the form 
\[
Q(\vx, \vy) =  Q_1(\vx)y_1 + \cdots + Q_m(\vx)y_m \, ,
\]
which \emph{explains} the received word $P$, i.e., for every $\va \in S^k$, $Q(\va, P(\va)) = 0$, and $Q$ satisfies some appropriate degree constraints. Here $m \leq s$ is a parameter. For technical reasons, we also end up imposing some more constraints on $Q$ in terms of its partial derivatives, the details of which can be found in \cref{sec: interpolation}. Each of these constraints can be viewed as a homogeneous linear equation in the coefficients of $Q$ over the field $\F(\vz)$. We choose the degree of $Q$ to be large enough to ensure that this system has more variables than constraints, and therefore, has a non-zero solution. 

This step is the interpolation step which shows up in any standard application of the polynomial method, and our set up is closest and a natural generalization of the set up in the list decoding algorithm of Guruswami and Wang \cite{GuruswamiW2013} for univariate multiplicity codes. 

The key property of the polynomial $Q$ thus obtained is that for every degree $d$ polynomial $f \in \F[\vx]$ whose encoding is close enough to $P$, 
\[
Q(\vx, \D(f)) = Q(\vx, \D_0(f), \D_1(f), \ldots, \D_{m-1}(f)) \equiv 0 \, .
\]
To see this, we note that from the upper bound on the degree of $Q$ and the fact that $f$ has degree at most $d$, the polynomial
$Q(\vx, \D(f)) \in \F(\vz)[\vx]$ is of not too high degree in $\vx$. Moreover, from the constraints imposed on $Q$ during interpolation, it follows that at every $\va \in S^k$ where the encoding of $f$ and $P$ agree, $Q(\vx, \D(f))$ vanishes with high multiplicity. Thus, if the parameters are favorably set, it follows that $Q(\vx, \D(f))$ has too many zeroes of high multiplicity on a grid, and hence by the multiplicity Schwartz-Zippel emma (see \cref{lem: PIL}), $Q(\vx, \D(f))$ must be identically zero.

We note that this is the only place in the proof where we use anything about the structure of the set of evaluation points, i.e., the set of evaluation points is a grid.  

\subsubsection*{Solving the equation to recover all close enough codewords}
As the final step of our algorithm,  we try to recover all polynomials $f \in \F[\vx]$ of degree at most $d$ such that 
\[
Q(\vx, \D(f)) = Q(\vx, \D_0(f), \D_1(f), \ldots, \D_{m-1}(f)) \equiv 0 \, .
\]
$Q(\vx, \D(f))$ can be viewed as a partial differential equation of order $m-1$ and degree one, and we construct all candidate solutions $f$ via the method of power series.  We start by trying all possible choices of field elements for coefficients  of monomials of degree at most $m-1$ in $f$, and iteratively recover the remaining coefficients of $f$ by reconstructing $f$ one homogeneous component at a time. Moreover, we observe that for each  choice of the initial coefficients, there is a unique lift to a degree $d$ polynomial. Thus, the number of solutions is upper bounded by the number of initial choices, which is at most $|\F|^{\binom{m+k-1}{k}}$. 

We note that this is one place where working with $\D_i(f)$ as opposed to having an equation in the individual partial derivatives of $f$ is of crucial help. Even though the equation $Q(\vx, \D(f)) = 0$ is a partial differential equation of high order in $f$, the fact that these derivatives appear in a structured form via the operators $\D_i(f)$ helps us prove a polynomial upper bound on the number of such solutions and  solve for $f$. Without this additional structure, it is unclear if one can prove a polynomial upper bound on the number of solutions of the corresponding equation. 

This reconstruction step is a multivariate generalization of similar reconstruction steps in the list decoding algorithms of Kopparty \cite{Kopparty2015} and Guruswami \& Wang \cite{GuruswamiW2013} for univariate multiplicity codes. Interestingly, this is also a special case of a similar reconstruction procedure in the work of Guo, Kumar, Saptharishi and Solomon \cite{GuoKSS2019}, where the polynomial $Q$ could potentially be of higher degree in $\vy$ variables, and is given to us via an arithmetic circuit of small size and degree and the goal is to show that all (low degree) polynomials $f$, satisfying $Q(\vx, \D(f)) \equiv 0$ have small circuits. 
In contrast, we are working with $Q$ which is linear in $\vy$ and we have access to the coefficient representation of this polynomial, and construct the solutions $f$ in the monomial representation. As a consequence, the details of this step are much simpler here, when compared to that in \cite{GuoKSS2019}. 

In this step of our algorithm viewing the encoding in terms of the differential operators $\D_i()$ turns out to be useful. The iterative reconstruction outlined above crucially uses the fact that for any homogeneous polynomial $g \in \F[\vx]$ of degree $r$, $\D_i(g)$ is a homogeneous polynomial in the $\vx$ variables of degree exactly  $r-i+1$. The other property that we use from $\D_i()$ is that given $\D_i(g)$ for any homogeneous polynomial $g$, we can uniquely read off all the partial derivatives of order $i-1$ of $g$, and via a folklore observation of Euler, uniquely reconstruct the polynomial $g$ itself (see \cref{lem: useful props of delta}). 

Finally, we note that the precise way of \emph{gluing} together the partial derivatives of order $i$  in the definition of the operator $\D_i()$ is not absolutely crucial here, and as is evident in \cref{lem: useful props of delta}, many other candidates would have satisfied the necessary properties. 

The details of this step are in \cref{sec: solving the equation}, and essentially complete the proof of \cref{thm:main}.  

\subsection{\cref{thm: constant list size}: Reducing the list size to a constant}

In \cref{sec: constant list size}, we combine our proof of \cref{thm:main} with the techniques in the recent work of Kopparty, Ron-Zewi, Saraf and Wootters \cite{KoppartyRSW2018} to show that the list size in the decoding algorithm in \cref{thm:main} can be reduced to a constant. 

The key to this step is the observation that since $Q(\vx, \vy)$ is linear in the $\vy$ variables, the solutions $f$ of the equation $Q(\vx, \D(f)) \equiv 0$ form an affine subspace of polynomials. The reconstruction algorithm in \cref{sec: solving the equation} in fact gives us an affine subspace $ V \subseteq \F[\vx]$ of polynomials of degree at most $d$ which consists of all the solutions of $Q(\vx, \D(f)) \equiv 0$. 

This is precisely the setting in the work of Kopparty, Ron-Zewi, Saraf and Wootters~\cite{KoppartyRSW2018} in the context of folded Reed-Solomon codes and univariate multiplicity codes, and we essentially apply their ideas off the shelf, and combine them with our proof of \cref{thm:main} to reduce the list size to a constant. 

In general, this idea of solving $Q(\vx, \D(f)) \equiv 0$ to recover a subspace, and then using the ideas in \cite{KoppartyRSW2018} to recover codewords in the subspace which are close to the received word has the added advantage that it can be applied over all fields. As an immediate consequence, we get an analog of \cref{thm:main} over infinite fields like rationals as well.

\section{Preliminaries}\label{sec: prelims}

\subsection{Notation}
We use the following notation. 
\begin{itemize}
\item $\F$ is the field we work over, and we assume the characteristic of $\F$ to be either zero or larger than the degree parameter $d$ of the message space. 
\item We use bold letters to denote tuples of variables (i.e., $\vecx$, $\vecz$, $\vy$ for $(x_1,\dots,x_k), (z_1,\dots,z_k)$ and $(y_1,\dots,y_m)$ respectively).  
\item We work with polynomials which are in general members of $\F(\vz)[\vx,\vy]$. We denote monomials in $\vx$ and $\vz$ by $\vx^{\ve}$ ($=\prod_{i \in [k]}x_i^{e_i}$), $\vz^{\ve}$ ($=\prod_{i \in [k]}z_i^{e_i}$) respectively where $\ve\in \Z_{\geq 0}^k$. The degree of the monomial is $\|\ve\|_1=\sum_{i=1}^{k}e_i$. 
\item For $\ve,\ve' \in \Z_{\geq 0}^{k}$ we say $\ve'\leq \ve$ iff for all $i \in [k]$ we have $e'_i\leq e_i$. Also, we use ${\ve \choose \ve'}$ to denote $\prod_{i\in [k]}\binom{e_i}{e'_i}$.
\item For a natural number $n$, $[n]$ denotes the set $\{1, 2, \ldots, n\}$. 
\end{itemize}

\subsection{Hasse derivatives}
Throughout the paper we work with Hasse derivatives: we interchangeably use the term partial derivatives.

\begin{definition}[Hasse Derivative]
\label{def:Hasser_Derv}
For a polynomial $f\in \F[\vx]$ the Hasse derivative of type $\ve$ is the coefficient of $\vz^{\ve}$ in the polynomial $f(\vx+\vz)\in \F[\vx,\vz]$.
We denote this by $\frac{\hpartial f}{\hpartial \vx^{\ve}}$ or $\frac{\hpartial f(\vx)}{\hpartial \vx^{\ve}}$
\end{definition}

We state some basic properties of Hasse Derivatives below. Some of these are taken from~\cite[Proposition~4]{DvirKSS2013}.
\begin{proposition}[Basic Properties of Hasse Derivatives]
\label{prop:Prop_of_Hasse_Derv}
Let $f,g\in \F[\vx]$ and consider $\ve,\ve'\in \Z_{\geq 0}^{k}$.
\begin{enumerate}
    \item  $\frac{\hpartial f}{\hpartial \vx^{\ve}}+\frac{\hpartial g}{\hpartial \vx^{\ve}}=\frac{\hpartial (f+g)}{\hpartial \vx^{\ve}}$.\label{item:hasse-1}
    \item If $f$ is a homogeneous polynomial of degree $d$ then $\frac{\hpartial f}{\hpartial \vx^{\ve}}$ is homogeneous polynomial of degree $d-\|\ve\|_1$.\label{item:hasse-2}
    \item If $f=\vx^{\ve'}$ then $\frac{\hpartial f}{\hpartial \vx^{\ve}}= \binom{\ve}{\ve'}\vx^{\ve-\ve'}$.\label{item:hasse-3}
    \item Hasse derivatives compose in the following manner:
    \[\frac{ \hpartial }{\hpartial
    \vx^\ve} \frac{\hpartial f(\vx)}{\hpartial \vx^{\ve'}} = {\ve + \ve'\choose \ve}\cdot \frac{\hpartial f(\vx)}{\hpartial \vx^{\ve+\ve'}}.\]\label{item:hasse-4}
    \item Product rule for Hasse derivatives: \[\frac{\hpartial \left(\prod_{i\in [w]}f_i\right) }{\hpartial \vx^{\ve}}=\sum_{\vu_1+\vu_2+\ldots+\vu_w=\ve}\left(\prod_{i\in [w]}\frac{\hpartial f_i}{\hpartial \vx^{\vu_i}}\right).\]\label{item:hasse-5}

\end{enumerate}
\end{proposition}
\begin{proof}
\cref{item:hasse-1,item:hasse-2,item:hasse-3,item:hasse-5} follow directly from \cref{def:Hasser_Derv}. For \cref{item:hasse-4}, observe that by linearity of Hasse derivatives we may assume WLOG that $f$ is a monomial, say $\vx^{\tilde{\ve}}$: in this case the claim follows from \cref{item:hasse-3} and the fact that $\binom{\tilde{\ve}}{\ve}\cdot \binom{\tilde{\ve}-\ve}{\ve'}= \binom{\ve + {\ve'}}{ \ve}\cdot\binom{\tilde{\ve}}{\ve+\ve'}$.
\end{proof}

\subsection{Multiplicity code}

We now define the notion of multiplicity of a polynomial $f\in \F[\vx]$ at a point $\va \in \F^{k}$. The multiplicity of $f$ at the origin is $\ell$ iff $\ell$ is the highest integer such that no monomial of total degree less than $\ell$ appears in the coefficient representation of $f$. We formalize this below using Hasse derivatives. 

\begin{definition}[multiplicity]
\label{def:Multiplicity}
A polynomial $f\in \F[\vx]$ is said to have multiplicity $\ell$ at a point $\va \in \F^{k}$, denoted by $\mult(f, \va)$, iff $\ell$ is the largest integer such that for all $\ve \in \Z_{\geq 0}^{k}$ with $\|\ve\|_1<\ell$ we have $\frac{\hpartial f}{\hpartial \vx^{\ve}}(\va)=0$. If no such $\ell$ exists then $\mult(f, \va)=\infty$.
\end{definition}

Dvir, Kopparty, Saraf and Sudan proved the following higher order multiplicity version  of the classical Schwartz-Zippel lemma. 
 \begin{lemma}[multiplicity SZ lemma {\cite[Lemma~2.7]{DvirKSS2013}}]\label{lem: PIL}
Let $\F$ be any field and let $S$ be an arbitrary subset of $\F$. Then, for any non-zero $k$-variate polynomial $P$ of degree at most $d$, 
\[
\sum_{\va \in S^k} \mult(P, \va) \leq d|S|^{k-1} \, .
\]
\end{lemma} 
The above lemma implies the classical SZ lemma, which states that two distinct $k$-variate polynomials of degree $d$ cannot agree everywhere on a grid $S^k$ for any set $S$ of size larger than $d$ trivially. This in particular tells us that the grid $S^k$ serves as hitting set for polynomials of degree at most $d$ provided $d < |S|$.  

As mentioned before, a multiplicity code over a grid $S^{k}$ consists of evaluations of the message polynomial $f$ along with its derivatives of various orders (up to $s-1$), at the points of the grid.
\begin{definition}[multiplicity code]
  \label{def:Mult_code}
  Let $s, k \in \N$, $d \in \Z_{\geq 0}$, $\F$ a field and $S \subset \F$  a non-empty finite subset. The $k$-variate order-$s$ multiplicity code of 
  degree-$d$ polynomials over $\F$ on the grid $S^k$ is defined as follows.

  Let $E := \{ \ve \in \Z_{\geq 0}^k \mid 0 \leq \|\ve \|_1 < s\}$. Note that $|E| = {s+ k -1 \choose k}$. The code is over alphabet $\F^E$ and has length $S^k$ (where the coordinates are indexed by elements of $S^k$).

  The code is an $\F$-linear map from  the space of degree $d$ polynomials in $\F[\vx]$ to $\left(\F^E\right)^{S^k}$. 
The encoding of $f\in \F[\vx]$ at a point $\va \in S^k$ is given by:
\[
  \enc_{s, S}(f)|_{\va} = \left(\frac{\hpartial f}{\hpartial \vx^\ve}(\va) : \ve \in E\right). \qedhere
 \] 
\end{definition}
\begin{remark}
  ~\\
  \begin{itemize}
    \item The distance of the code is exactly $\delta :=  1 - \frac{d}{s|S|}$ and the rate of the of the code is $\frac{{d+k \choose k}}{{s+ k -1 \choose k} \cdot |S|^k}$. 
    \item As mentioned in the introduction we can also view the encoding by clubbing partial derivatives of the same degree. Thus, the encoding of $f$ at a point $\va$ is 
    $(\D_0(f)(\va), \D_1(f)(\va), \ldots, \D_{s-1}(f)(\va))\in \F[\vz]^s$ where $\D_i(f)(\va) = \sum_{\ve: \|\ve\|_1= i} \vz^{\ve} \cdot \frac{\hpartial f(\vx)}{\hpartial \vx^{\ve}}(\va)$.
    \item We think of $k$, $m$ and $s$ as constants, but $m$ much larger than $k$ and $s$ is much larger than $m$. The precise trade-offs will be alluded to when we need to set parameters in our proofs.  
\end{itemize}
\end{remark}


\subsection{Computing over polynomial rings}

In this section, we state a few basic results that show how to perform algebraic operations over polynomial rings.

The following lemma, proved via an easy application of polynomial interpolation, lets us construct the coefficient representation of a polynomial given an arithmetic circuit for it. 
\begin{lemma}\label{lem: coefficient representation from circuit}
  Let $k \in \N$. There exists a deterministic algorithm that takes as input an arithmetic circuit $C$ of size $s$ that computes a $k$-variate polynomial $P \in \F[\vz]$ of degree at most $d$ and outputs the coefficient vector of $P$ in at most $\poly(d^k, s)$ field operations over $\F$
\end{lemma}
\begin{proof}
From \cref{lem: PIL}, we know that no two degree $d$ polynomials can agree everywhere on a grid of size larger than $d$. So, we pick an arbitrary subset $S$ of $\F$ of size $d+1$ and evaluate the circuit $C$ at all points on the grid $|S|^k$. This requires at most $\poly(d^k, s)$ field operations. Now, given these evaluations, we set up a linear system in the coefficients of $P$ where for every $\va$ in the grid, we have a constraint of the form $P(\va) = C(\va)$. We know that this system has a solution. Furthermore, from \cref{lem: PIL}, we know that this system has a unique solution. 

Solving this system gives us the coefficient vector of $P$ and requires at most $d^k$ additional field operations. 
\end{proof}

The next lemma tells us how to perform linear algebra over the polynomial ring $\F[\vz]$.
 \begin{lemma}[linear algebra over polynomial rings]\label{lem: lin algebra}
Let $A(\vz) \in \F[\vz]^{t' \times t}$ be a matrix such that each of its entries is a polynomial of degree at most $m$ in the variables $\vz = (z_1, z_2, \ldots, z_k)$ and $t' \leq t$. Then, there is a deterministic algorithm which takes as input the coefficient vectors of the entries of $A$ and outputs a non-zero vector $\vu \in \F[\vz]^{t}$ in time $\poly(m^k, t^k)$ such that $A\cdot \vu = \mathbf{0}$. Moreover, every entry in $\vu$ is a polynomial of degree at most $tm$.
\end{lemma}
\begin{proof}
As a first step, we reduce this to the problem of solving a linear system of the form $A'\cdot \vu' = \vb$, where $A'$ and $\vb$ have entries in $\F[\vz]$ of degree at most $m$, and $A'$ is a square matrix of dimension at most $t'$, which is non-singular. At this point, we can just apply Cramer's rule to find a solution of this system. 

Since $t' \leq t$, the rank $r$ of $A(\vz)$ over $\F(\vz)$ is at most $t'$. Thus, there is a square submatrix $A'(\vz)$ of $A$ such that $\det(A')$ is a non-zero polynomial of degree at most $mr \leq mt'$ in $\F[\vz]$. For a hitting set $H_{mt', k}$ of polynomials of degree at most $mt'$ on $k$ variables over $\F$, we consider the set of matrices $\{A(\vc) : \vc \in H_{mt', k}\}$. From the guarantees of the hitting set, we know that there is a $\vc \in H_{mt', k}$ such that $A'(\vc)$ is of rank equal to $r$. Let $\vc_0 \in H_{mt', k}$ be such that the rank of $A(\vc_0)$ over $\F$ is maximum among all matrices in the set $\{A(\vc) : \vc \in H_{mt', k}\}$. Moreover, let $A'(\vz)$ be a submatrix of $A(\vz)$ such that $\rank(A'(\vc_0))$ equals $\rank(A(\vc_0))$. From \cref{lem: PIL}, there is an explicit hitting set $H_{mt', k}$ of size at most $(mt' + 1)^k \leq (mt + 1)^k$. Thus, we can find $A'(\vz)$ of rank equal to the rank of $A(\vz)$ with at most  $\poly(m^k, t^k)$ field operations over $\F$. Without loss of generality, let us assume that $A'$ is the top left submatrix of $A$ of size $r$. Clearly, the $(r+1)$-{st} column of $A$ is linearly dependent on the first $r$ columns of $A$ over the field $\F(\vz)$.  In other words, the linear system given by 
\[
A'\cdot \vu' = \vb \, 
\] 
where $\vb = (A_{1, r+1}, A_{2, r+1}, \ldots, A_{r, r+1})$, has a solution in $\F(\vz)$. Moreover, for every solution $\vu'$ of this system, where $\vu' = (u_1', u_2', \ldots, u_r')$, the $t$ dimensional vector $(u_1', u_2', \ldots, u_r', -1, 0, \ldots, 0)$ is in the kernel of $A(\vz)$. Also, since $A\cdot \vu = \mathbf{0}$ is a homogeneous linear system, for any non-zero polynomial $P(\vz)$, $(P\cdot u_1', P\cdot u_2', \ldots, P\cdot u_r', -P, 0, \ldots, 0)$ continues to be a non-zero vector in the kernel of $A(\vz)$.

Since $A'$ is non-singular, $\vu' = (A')^{-1}\cdot \vb$ is a solution to this system. Moreover, by Cramer's rule, $(A')^{-1} = \textsf{adj}(A')/\det(A')$, where $\textsf{adj}(A')$ is the adjugate matrix of $A'$ and $\det(A')$ is its determinant. Since, every entry of $\textsf{adj}(A')$  is a polynomial in $\F[\vz]$ of degree at most $tm$, we get a solution of the form $\vu' = (p_1/\det(A'), p_2/\det(A'), \ldots, p_r/\det(A'))$ where each $p_i$ is a polynomial in $\F[\vz]$ of degree at most $tm$. By getting rid of the denominators by scaling by $\det(A')$, we get that the non-zero $t$ dimensional vector $(p_1, p_2, \ldots, p_r, -\det(A'), 0, \ldots, 0)$ is in the kernel of $A(\vz)$. 

Moreover, using the fact that the determinant polynomial has a polynomial size efficiently constructible circuit, and \cref{lem: coefficient representation from circuit}, we can output this vector, with each entry being a list of coefficients in $\F$ in time $\poly(m^k, t^k)$ via an efficient deterministic algorithm. 
 \end{proof}


\section{List decoding the multivariate multiplicity code}\label{sec:polylist}
In this section, we prove \cref{thm:main}. We follow the outline of the proof described in \cref{sec: proof overview}. We start with the interpolation step.
\subsection{Viewing the encoding as a formal power series}
The message space is the space of $k$-variate polynomials of degree at most $d$ over $\F$. In the standard encoding, we have access to evaluations of the polynomial and all its derivatives of order up to $s-1$ on all points on a grid  $S^k \subseteq \F^k$. 

For our proof, it will be helpful to group the derivatives of the same order together. 
\begin{definition}\label{defn: delta}
Let  $f \in \F[\vx]$ be a polynomial. Then, for any $i\in \Z_{\geq 0}$, 
$\D_{i}(f)$ is defined as 
\[
\D_i(f) := \sum_{\ve: \|\ve\|_1= i} \vz^{\ve} \cdot \frac{\hpartial f(\vx)}{\hpartial \vx^{\ve}} \, . \qedhere
\]
\end{definition}
So, we have a distinct monomial in $\vz$ attached to each of the derivatives. The precise form of the monomial in $\vz$ is not important, and all that we will use is that these monomials are linearly independent over the underlying field, don't have very high degree and there aren't too many variables in $\vz$. 

Now, we think of the encoding of $f$ as giving us the evaluation of the tuple of polynomials $\D(f) = \left(\D_0(f(\vx)), \D_1(f(\vx)), \ldots, \D_{s-1}(\vx)\right) \in \F(\vz)[\vx]^{s}$ as $\vx$ takes values in $\F^k$. 

Note that $\D_i(f)$ is a homogeneous polynomial of degree at equal to $i$ in $\vz$.  

\subsection{The $\tau$ operator}

We will need to compute the Hasse derivative of $\D_i(f)$ with
respect to $\vx^\ve$, i.e., $\frac{\hpartial \D_i(f)}{\hpartial
  \vx^{\ve}}$. From the definition of $\D_i(f)$, we have
\begin{align*}
  \frac{\hpartial \D_i(f)}{\hpartial
    \vx^{\ve}}
  & = \sum_{\ve': \|\ve'\|_1 =i} \vz^{\ve'} \cdot \frac{ \hpartial }{\hpartial
    \vx^\ve} \frac{\hpartial f(\vx)}{\hpartial \vx^{\ve'}} = \sum_{\ve':
    \|\ve'\|_1 =i} \vz^{\ve'} \cdot {\ve + \ve'\choose \ve}\cdot \frac{\hpartial f(\vx)}{\hpartial
    \vx^{\ve+\ve'}} \\
  &= \sum_{\ve':
    \|\ve'\|_1 =i} \vz^{\ve'} \cdot {\ve +\ve' \choose \ve} \cdot \coeff_{\vz^{\ve +
    \ve'}}(\D_{i+\|\ve\|_1} f(\vx))\, .
\end{align*}
The key point to note is that the Hasse derivative of $\D_i(f)$ with
respect to $\vx^\ve$ can be read off the coefficients of
$\D_{i+\|\ve\|_1}(f)$.

This motivates the following definition. Consider a tuple $P=(P_0, P_1, \ldots, P_{s-1})$, where for each $i$, $P_i$ is a homogeneous polynomial of degree $i$ in $\F[\vz]$. For any $\ve \in \Z_{\geq 0}^k$, and $i \leq s-1$ such that $i + \|\ve\|_1 \leq s-1$, we define 
\[
\tau_{\ve}^{(i)}(P) := \sum_{\ve': \|\ve'\|_1 = i} \vz^{\ve'}\cdot  {\ve +\ve' \choose \ve} \cdot \coeff_{\vz^{\ve + \ve'}}(P_{i + \|\ve\|_1})\, .
\]

Thus, for $\D(f) = \left(\D_0(f(\vx)), \D_1(f(\vx)), \ldots,
  \D_{s-1}(\vx)\right)$, we have
\[
  \tau_\ve^{(i)}(\D(f)) = \frac{\hpartial \D_i(f)}{\hpartial \vx^\ve}\, .
\]

\subsection{Interpolation step}\label{sec: interpolation}
Let $P$ be the received word, Thus, we are given a collection of $s$-tuples of polynomials $P(\va) = (P_0(\va), P_1(\va), \ldots, P_{s-1}(\va))$ for every $\va \in S^k$, where each $P_i(\va)$ is a homogeneous polynomial of degree $i$ in $\vz$. From the earlier definition of $\tau$, given such a $P(\va)$, we have $\tau^{(i)}_{\ve}(P(\va))$ for every $i \leq m$ and $\ve$ with $\|\ve\|_1 \leq s-1-m$. 
\begin{lemma}\label{lem: interpolation}
Let $k$ and $s$ be constants. For every natural number $m \leq s-1-k$, and $D = 10|S|(s-m)/m^{1/k}$, there is a non-zero polynomial $Q(\vx, \vy) =  Q_1(\vx)y_1 + \cdots + Q_m(\vx)y_m \in \F(\vz)[\vx, \vy]$ such that 
\begin{itemize}
\item For every $i \in \{1, 2, \ldots, m\}$, the $\vx$-degree of each $Q_i$ is at most $D$. 
\item For every $\va \in S^k$ and every $\ve \in \Z_{\geq 0}^k$ such that $0 \leq \|\ve\|_1 \leq s-1-m$, $\D_{\ve}(Q)(\va) = 0$, where
\[
\D_{\ve}(Q)(\va) :=  \sum_{i = 1}^m \sum_{\ve' \leq \ve}\frac{\hpartial Q_i(\vx)}{\hpartial \vx^{\ve'}}(\va)\cdot \tau^{(i-1)}_{\ve - \ve'}(P(\va)) \, .
\]
Here, $\ve' \leq \ve$ means that $\ve$ dominates $\ve'$ coordinate
wise.
\end{itemize}
Moreover, the coefficients of $Q$ are polynomials in $\F[\vz]$ of degree at most $O(|S|^ks^{2k})$, and such a $Q$ can be deterministically constructed by using at most $\poly(|S|^{k^2},s^{k^2}, d^k)$ operations over the field $\F$.  
\end{lemma}
\begin{proof}
We start by showing the existence of a polynomial $Q$ with the appropriate degree constraints, followed by an analysis of the running time. 

\paragraph*{Existence of $Q$. }
We view the above constraints as a system of linear equations over the field $\F(\vz)$, where the variables are the coefficients of $Q$. The number of homogeneous linear constraints is $|S|^k\binom{s - m + k}{k}$ and the number of variables is $m\binom{D + k}{k}$. 

By using the fact that $k$ is much smaller than $s$, and a crude approximation of the binomial coefficients, we have $|S|^k\binom{s - m + k}{k} \leq \left(2e|S|(s-m)/k\right)^k$ and $ m\binom{D + k}{k} > m(D/k)^k$. Plugging in the value of $D$, we get $m(D/k)^k = \left(10|S|(s-m)/k\right)^k$, which is clearly greater than the number of constraints. Hence, there is a non-zero solution, where the coefficients of the polynomial are from the field $\F(\vz)$, i.e., are rational functions in $\vz$.

Next we analyze the degree of these coefficients and show that we can recover such a $Q$ efficiently, with the appropriate degree bounds. 

\paragraph*{The running time. }
For the running time, we recall that each $\tau_\ve^i$ is a polynomial of degree at most $m-1$ in the $\vz$ variables. As a consequence, observe that the linear system we have for the coefficients of $Q$ is of the form $A\cdot \vu = 0$, where $A$ is a matrix with dimension at most $O(|S|^k(s-m)^{k})$ over the ring $\F[\vz]$, and every entry of $A$ is a polynomial in $\F[\vz]$ of degree at most $m$. From \cref{lem: lin algebra}, we get that we can find a non-zero solution in $\F[\vz]$ using at most $\poly(|S|^{k^2}, s^{k^2})$ field operations over $\F$. Moreover, each of the coordinates of this output vector is a polynomial of degree at most $O(|S|^k(s-m)^{k})\cdot m =O(|S|^{k}s^{2k})$ in $\F[\vz]$. 
\end{proof}

Going forward, we work with the polynomial $Q$ and the degree parameter $D$ as set in   \cref{lem: interpolation}. 
\subsection{Close enough codewords satisfy the equation}
We now show that  for every polynomial $f \in \F[\vx]$ of degree at most $d$ whose encoding is close enough to the received word $P$, $f$ \emph{satisfies} the equation $Q$ in some sense. 
\begin{lemma}\label{lem: close enough code words satisfy the equation}
 If $f \in \F[\vx]$ is a degree $d$ polynomial such that the number of $\va \in S^k$ which satisfy
\[
P(\va) = \D(f)(\va) \, ,
\]
is at least $T > (D + d)\cdot |S|^{k-1}/(s-m)$, then $Q(\vx, \D_0(f), \D_1(f), \ldots, \D_{m-1}(f))$ is identically zero as a polynomial in $\F(\vz)[\vx]$. 
\end{lemma}

\begin{proof} Define the polynomial $R \in \F(\vz)[\vx]$ as follows
\[R(\vx) := Q(\vx, \D_0(f), \D_1(f), \ldots, \D_{m-1}(f)) = 
  \sum_{i=1}^m Q_i(\vx) \cdot \D_{i-1}(f)\, .\]
$R$ is a polynomial in $\vx$ of degree at most $D + d$ over the field
$\F(\vz)$. Whenever $\va$ satisfies that $P(\va) = \D(f)(\va)$, from the definitions of $\tau_\ve^{(i)}$ and $\D_\ve$, we have that for
all $\ve$ such that $0 \leq \|\ve \|_1 \leq s-m-1$, 
\begin{align*}
  \frac{\hpartial R(\vx)}{\hpartial \vx^\ve}(\va)
  & =  \sum_{i =
    1}^m \sum_{\ve' \leq \ve} \frac{\hpartial Q_i(\vx)}{\hpartial
    \vx^{\ve'}}(\va)\cdot \frac{\hpartial \D_{i-1}(f)}{\hpartial
    \vx^{\ve-\ve'}}(\va) \\
  & =  \sum_{i =
    1}^m \sum_{\ve' \leq \ve} \frac{\hpartial Q_i(\vx)}{\hpartial
    \vx^{\ve'}}(\va)\cdot \tau^{(i-1)}_{\ve - \ve'}(P(\va))\\
  &= \Delta_\ve(Q) (\va) \\
  &= 0 \, .
\end{align*}  
Hence, at every point of agreement between $\D(f)$ and the received
word $P$, $R(\vx)$ vanishes with multiplicity at least $s-m$. From \cref{lem: PIL}, we know that if
\[
T(s-m) > (D + d)|S|^{k-1} \, ,
\]
then, $R$ must be identically zero. 
\end{proof}
Let us try to get a sense of the parameters here. The relative distance of this code is $\delta = 1-\frac{d}{s|S|}$. Now, in $\frac{T}{|S|^k} > \frac{D + d}{|S|(s-m)}$, plugging in the value of $D$ from the earlier discussion gives us 
\begin{align*}
\frac{T}{|S|^k} &>\frac{d}{|S|(s-m)} + \frac{10|S|(s-m)/m^{1/k}}{|S|(s-m)} \\
 &=\frac{10}{m^{1/k}} + \left(\frac{s}{s-m}\right)\cdot \frac{d}{s|S|} \\
 &= \frac{10}{m^{1/k}} + \left(\frac{m}{s-m}\right)\cdot \frac{d}{s|S|} + \frac{d}{s|S|} \, .
\end{align*}
In our final analysis for the proof of \cref{thm:main}, we choose $m$ and $s$ large enough as a function of $\epsilon$, so that this bound is of the form $\epsilon + (1 - \delta)$, which is precisely what is claimed in \cref{thm:main}. 

\subsection{Solving the equation to find close enough codewords}\label{sec: solving the equation}
All that remains now is to solve equations of the form $Q(\vx, \D_0(f), \D_1(f), \ldots, \D_{m-1}(f))$ to recover $f$. This would be done via iteratively constructing $f$ one homogeneous component at a time. We will need the following easy observations. 

\begin{lemma}\label{lem: useful props of delta}
Let $\F$ be a field of characteristic zero or larger than $d$. Let $f\in \F[\vz]$ be a polynomial of degree $d$, and for every $i\in \Z_{\geq 0}$, $\D_i$ be the differential form of  order $i$ as defined in \cref{defn: delta}. Then, the following are true.  
\begin{itemize}
\item For each $i \in \Z_{\geq 0}$, $\D_i(f)$ is homogeneous in $\vz$ and has degree $i$ in the $\vz$ variables. Moreover, for any monomial $\vz^{\ve}$ of degree $i$, its coefficient in $\D_i(f)$ equals $\frac{\hpartial f}{\hpartial \vx^{\ve}}$. 
\item If $f$ is a homogeneous polynomial, then, for every $i \leq d$, $f$ can be uniquely recovered from all its partial derivatives of order $i$. As a consequence, for any homogeneous $f$, given the formal polynomial $\D_i(f)$,  we can recover $f$. 
\end{itemize}
\end{lemma}
\begin{proof}
The first item follows directly from the definition of $\D$ in \cref{defn: delta}. 

The second item follows from an immediate generalization of the following well known observation of Euler to Hasse derivatives. For any homogeneous polynomial $f$ of degree $d$, 
\[
d\cdot f = \sum_{i} x_i\cdot \frac{\hpartial f(\vx)}{\hpartial x_i} \, .
\]

We also have that 
\[
\frac{ \hpartial }{\hpartial
    \vx^\ve} \frac{\hpartial f(\vx)}{\hpartial \vx^{\ve'}} = {\ve + \ve'\choose \ve}\cdot \frac{\hpartial f(\vx)}{\hpartial \vx^{\ve+\ve'}}\, .
\]
Using this we can compute the first order Hasse derivatives of $\frac{\hpartial f}{\hpartial \vx^{\ve'}}$ for all $\|\ve'\|_1=i-1$ from $\D_i(f)$.
So, for any $i$, given all Hasse derivatives of degree $i$, we can recover Hasse derivatives of degree $i-1$ (using Euler's formula), and so on, till we recover $f$. 
\end{proof}

\begin{remark}
We remark that the second item in \cref{lem: useful props of delta} is false for fields of small characteristic. For instance, if the characteristic is smaller than $d$, then even for a non-zero $f$, all its first order derivatives could be zero, and hence $f$ cannot be recovered from its first order derivatives. 
\end{remark}

The following lemma shows that under very mild conditions on $Q(\vx, \vy)$, we can (efficiently) recover all polynomials $f$ of degree at most $d$ such that $Q(\vx, \D_{< m}(f))\equiv 0$. This will complete all the ingredients needed for the proof of \cref{thm:main}. 

\begin{lemma}\label{lem: reconstruction}
Let $\F$ be a finite field of characteristic larger than $d$ and let $Q(\vx,\vy) =  Q_1y_1 + \cdots + Q_{m}y_m$ be any non-zero polynomial in $\F[\vz, \vx, \vy]$ where, $\deg_{\vx}(Q) \leq D + d$, $\deg_{\vz}(Q) \leq \Gamma$ and $Q_i$ does not depend on $\vy$. Then, there is a deterministic algorithm that outputs all polynomials $f \in \F[\vx]$ of degree at most $d$ such that 
\[
Q(\vx, \D_0(f),\D_1(f), \ldots, \D_{m-1}(f)) \equiv 0 \, .
\]

Moreover, the algorithm requires at most $\poly \left(D^{k}, d^k, |\F|^{\binom{m+k}{m}}, \Gamma^k \right)$ arithmetic operations over the underlying field $\F$. 
\end{lemma}
\begin{proof}
We will reconstruct $f$ iteratively,  one homogeneous component at a time. This iterative process has to be started by fixing the homogeneous components of $f$ of degree at most $m$, and as will be evident from the discussion ahead, every fixing of this initial seed can be lifted to a unique $f$ of degree at most $d$ satisfying  
\[
Q(\vx,\D_0(f),\D_1(f), \ldots, \D_{m-1}(f)) \equiv 0 \, .
\]

Before starting the reconstruction, we  need to ensure appropriate non-degeneracy conditions which are typical in iterative reconstruction arguments of this kind. 

\paragraph*{Preprocessing. }
We know from the hypothesis of the lemma that $Q$ depends on at least one $y$ variable. Let $j$ be the largest index in $\{1, \ldots, m\}$ such that $Q$ depends on $y_j$, i.e., $Q_{j}$ is non-zero and $Q_i$ is identically zero for all $i > j$. For the ease of notation, we  shall assume that $j = m$, thus, $Q_m$ is a non-zero polynomial. Recall that $f$ is a polynomial in $\F[\vx]$ and each $\D_i(f)$ is a polynomial in $\F[\vx, \vz]$. 
 
Since $Q_m(\vx) \in \F[\vx]$ is a non-zero polynomial, there is an $\va \in \F^k$ such that $Q_m(\va)\neq 0$.\footnote{This is assuming $\F$ is large enough, else we can find such an $\va$ in a large enough extension field of $\F$.} Replacing the variable $x_i$ by $x_i' + a_i$ (i.e., translating the origin), we can ensure that in this translated coordinate system, $Q_m(\vx'+ \va)$ is non-zero at the origin, i.e., when $\vx'$ is set to $\mathbf{0}$. We work in this translated coordinate system for the ease of notation. Observe that every solution $f(\vx) \in \F[\vx]$ is bijectively mapped to a solution $\tilde{f}(\vx') = f(\vx' + \va) \in \F[\vx']$ and given $\tilde{f}$, we can efficiently recover $f$. Also, note that $\D_i(f)(\vx' + \va) = \D_i(f(\vx' + \va)) = \D_i(\tilde{f}(\vx))$, i.e., taking derivatives and then setting $\vx = \vx' + \va$ is equivalent to first doing the translation $\vx = \vx' + \va$ and then taking derivatives. Let 
\[
Q'(\vx') := Q(\vx'+ \va) = Q_1(\vx'+ \va)y_1 + \cdots + Q_{m}(\vx'+ \va)y_m \, , 
\]
and let  $I = \langle x_1', \ldots, x_k' \rangle$ be the ideal generated by $\{x_1', \ldots, x_k'\}$. 


\paragraph*{Iterative Reconstruction. }
We are now ready to describe the iterative reconstruction of $\tilde{f}$. 
\begin{itemize}
\item {\textbf{Base Case :}} We will try all possible values for the coefficients of monomials of degree at most $m$ in $\tilde{f}$ from the field $\F$. There are $|{\F}|^{\binom{m + k}{k}}$ possible choices. The next steps are going to uniquely lift each of these candidate solutions to a degree $d$ polynomial, so the number of solutions  remains $|{\F}|^{\binom{m + k}{k}}$. 
\item \textbf{Induction Step :} We now assume that we have recovered
  $\tilde{f}_0, \tilde{f}_1, \ldots, \tilde{f}_t \in \F[\vx']$ for some $t \geq m$, where $\tilde{f}_i$ is a homogeneous component of
  $\tilde{f}$ of degree $i$. The goal is to recover $\tilde{f}_{t+1}$, the $(t+1)$-st
  homogeneous component. Let $\tilde{f}_{\leq t} = \tilde{f}_0 + \tilde{f}_1 + \cdots + \tilde{f}_{t}$. Now,  let us consider the equation $Q'(\vx',\D_0(\tilde{f}), \D_1(\tilde{f}), \ldots, \D_{m-1}(\tilde{f})) = 0$ when we work modulo the ideal $I^{t - m + 3}$. Clearly, the homogeneous components of $\tilde{f}$ of degree larger than $t+1$ do not contribute anything modulo $I^{t - m + 3}$, and so we have, 
\[
Q'(\vx', \D_0(\tilde{f}_{\leq t}), \D_1(\tilde{f}_{\leq t}), \ldots, \D_{m-1}(\tilde{f}_{\leq t} + \tilde{f}_{t+1})) = 0 \mod I^{t - m + 3} \, .
\]
Using the linearity of $\D_i$ and the fact that $Q'$ is linear in $\vy$, we get 
\[
Q'(\vx',\D_0(\tilde{f}_{\leq t}), \D_1(\tilde{f}_{\leq t}), \ldots, \D_{m-1}(\tilde{f}_{\leq t})) + Q_m(\vx'+ \va)\cdot \D_{m-1}(\tilde{f}_{t+1})= 0 \mod I^{t - m + 3} \, .
\]
We know that the degree of $\D_{m-1}(\tilde{f}_{t+1})$ equals $t+1 - (m-1) = t - m + 2$, and it is homogeneous in $\vx'$. Also, we have ensured in the preprocessing phase that $Q_m(\vx'+ \va) \mod I = Q_m(\va)$ is some non-zero constant $\F$. Thus, this is a non-trivial linear equation in $\D_{m-1}(\tilde{f}_{t+1})$ and if we can use it to recover all the partial derivatives of $\tilde{f}_{t+1}$ of order $m-1$, we can then use \cref{lem: useful props of delta} to recover $\tilde{f}_{t+1}$ itself. We elaborate on the details of this step of recovering the partial derivatives of $\tilde{f}_{t+1}$ from $\D_{m-1}(f_{t+1})$ next. 
\end{itemize}  

\paragraph*{Recovering partial derivatives of $\tilde{f}_{t+1}$ from $\D_{m-1}(\tilde{f}_{t+1})$. } Recall that since $\tilde{f}_{t+1}$ is a homogeneous polynomial in $\F[\vx]$ of degree $t+1$, each of its partial derivatives of order $m-1$ is a homogeneous polynomial in $\vx'$ of degree equal to $t+1 - (m-1) = t-m + 2$. Thus, 
\[
\D_{m-1}(\tilde{f}_{t+1}) := \sum_{\ve: \|\ve\|_1= m-1} \vz^{\ve} \cdot \frac{\hpartial \tilde{f}_{t+1}(\vx')}{\hpartial \vx'^{\ve}} \, .
\]
is a homogeneous polynomial in both $\vz$ and $\vx'$, with degree $m-1$ in $\vz$ and degree $t-m+2$ in $\vx'$. Our goal is to recover the coefficients of all monomials $\vz^{\ve}$ of degree $m-1$ in $\vz$ when viewing $\D_{m-1}(\tilde{f}_{t+1})$ as a polynomial in $\F[\vx][\vz]$, and we have access to  the expression 
\[
Q'(\vx', \D_0(\tilde{f}_{\leq t}), \D_1(\tilde{f}_{\leq t}), \ldots, \D_{m-1}(\tilde{f}_{\leq t})) =  - Q_m(\va)\D_{m-1}(\tilde{f}_{t+1})  \mod I^{t - m + 3}  \, .
\]
As a first step, observe that the polynomial $Q_m(\va)\D_{m-1}(\tilde{f}_{t+1})$ has degree at most $\Gamma + m-1$ in $\vz$ and degree exactly $t-m+2$ in $\vx'$. Moreover, since $Q_m(\va) \in \F[\vz]$ is non-zero, the polynomials $\{Q_m(\va)\vz^{\ve}: \deg(\vz^{\ve}) = m-1\}$ are linearly independent as polynomials of degree at most $\Gamma + (m-1)$ in $\vz$ over the field  over the field $\F$. Therefore, for any hitting set $H \subseteq \F^k$  for $k$-variate polynomials of degree at most $\Gamma + (m-1)$, the evaluation vectors $\eval_{H}(Q_m(\va)\vz^{\ve})$ of these polynomials on $H$ are linearly independent over $\F$. So, for every $\vx'^{\ve_0}$ of degree $m-1$, there exists an $\F$ linear combination of the polynomials $\{Q_m(\va)\D_{m-1}(\tilde{f}_{t+1}){\vb} : \vb \in H\}$ which equals $\frac{\hpartial \tilde{f}_{t+1}(\vx')}{\hpartial \vx'^{\ve}}$. Moreover, such a linear combination can be found (e.g. via Gaussian Elimination over the field $\F$) efficiently in the size of this linear system. 

Thus, to recover the partial derivatives of order $m-1$ of $\tilde{f}_{t+1}$ given a monomial representation of $Q_m(\va)\D_{m-1}(\tilde{f}_{t+1})$, we consider the hitting set $H$ of size $O(\Gamma\cdot m)^k$ for $k$-variate degree $\Gamma(m-1)$ polynomials given by \cref{lem: PIL}, compute the evaluation of the polynomials  
\[
Q_m(\va)\cdot \D_{m-1}(\tilde{f}_{t+1}) = \sum_{\ve: \|\ve\|_1= m-1} Q_m(\va)\vz^{\ve} \cdot \frac{\hpartial \tilde{f}_{t+1}(\vx)}{\hpartial \vx^{\ve}} \, ,
\]
at every $\vb \in H$, and take appropriate weighted linear combinations to recover each of the partial derivatives $\frac{\hpartial \tilde{f}_{t+1}(\vx)}{\hpartial \vx^{\ve}}$. 

Since $Q'(\vx', \D_0(\tilde{f}_{\leq t}), \D_1(\tilde{f}_{\leq t}), \ldots, \D_{m-1}(\tilde{f}_{\leq t})) $ is a polynomial of degree at most $\Gamma + m$ in $\vz$ and at most $D+d$ in $\vx$, we can do the evaluations by writing the coefficient vector of this polynomial in time $\poly(D^k, d^k, \Gamma^k, m^k)$ and doing evaluations one monomial at a time.

\paragraph*{The running time. }  Observe that we can go from the original polynomial $Q$ to the polynomial $Q'$ by finding an appropriate $\va$ deterministically in time at most $(D+d)^k$ by just querying all points on a large enough grid in $\F^k$ (or a grid in an extension field of $\F$, if $\F$ isn't large enough). This follows from \cref{lem: PIL}. 
 
 Once we have $Q'$, we reconstruct $f$ in $d$ iterations, so it suffices to estimate the cost of each iteration. As we just argued in the earlier part of the proof, every iteration just involves evaluating the polynomial $Q'(\vx', \D_0(\tilde{f}_{\leq t}), \D_1(\tilde{f}_{\leq t}), \ldots, \D_{m-1}(\tilde{f}_{\leq t})) $ at a hitting set $H$ of size at most $\poly(\Gamma^k, m^k)$ and solving about $m^k$ linear systems of the same size. The straightforward implementation of this takes no more than $\poly(D^k, d^k, \Gamma^k, m^k)$ field operations. 
\end{proof}


As is evident from the proof of \cref{lem: reconstruction}, the following more structured version of \cref{lem: reconstruction} is true. 
\begin{lemma}\label{lem: subspace reconstruction}
Let $\F$ be a field of characteristic zero or larger than $d$ and let $Q(\vx,\vy) = Q_1y_1 + \cdots + Q_{m}y_m$ be any polynomial in $\F[\vz, \vx, \vy]$ where, $\deg_{\vx}(Q) \leq D + d$, $\deg_{\vz}(Q) \leq \Gamma$ and $Q_i$'s do not depend on $\vy$. Then, there is a deterministic algorithm that outputs a linear space of polynomials in $\F[\vx]$ of dimension at most $\binom{m+k}{k}$ over $\F$ which contains all polynomials $f \in \F[\vx]$ of degree at most $d$ such that 
\[
Q(\vx, \D_0(f),\D_1(f), \ldots, \D_{m-1}(f)) \equiv 0 \, .
\]

Moreover, the algorithm requires at most $\poly \left(D^{k}, d^k, \Gamma^k \right)$ arithmetic operations over the underlying field $\F$. 
\end{lemma}
To bound the true running time  of the algorithm in \cref{lem: subspace reconstruction}, we need to add a $\poly(\log \F)$ factor in the the upper bound on the field operations for finite fields and a polynomial factor in the bit complexity of the input over the field of rational numbers. While working over rationals, we might need a bit more care to solve the linear systems appearing in the proof of \cref{lem: reconstruction} efficiently, since the naive implementation of Gaussian Elimination might blow up the bit complexity of the numbers appearing at various intermediate stages. 


\subsection{Putting things together}
We are now ready to prove \cref{thm:main}.
\begin{proof}[Proof of \cref{thm:main}]
We start by setting the parameters. $\epsilon$ and $k$ are fixed apriori, and we choose $s, m$ such that $s = m^2$ and $m$ is large enough so that 
\[
\frac{10}{m^{1/k}} + \frac{m}{s-m} < \epsilon \ .
\]
With this choice of parameters, we use \cref{lem: interpolation} to construct a non-zero polynomial $Q$ which \emph{explains} the received word $P$.  Then, we use \cref{lem: reconstruction} to find all polynomials $f\in \F[\vx]$ of degree at most $d$ such that 
\[
Q(\vx, \D_0(f), \D_1(f), \ldots, \D_{m-1}(f)) \equiv 0 \, .
\]
We know, from \cref{lem: reconstruction} that  the number of such solutions is upper bounded by $|\F|^{\binom{m+k}{k}}$ and from \cref{lem: close enough code words satisfy the equation} that every polynomial $f$ of degree at most $d$ in $\F[\vx]$ such that $\dist(\enc(f), P)$ is at most $(1-\delta) - \epsilon$, where $\delta = 1-d/(s|S|)$ satisfies the equation
\[
Q(\vx, \D_0(f), \D_1(f), \ldots, \D_{m-1}(f)) \equiv 0 \, .
\]
Thus, all such polynomials $f$ are included in the list of outputs. The running time of the algorithm immediately follows from the running time guarantees in \cref{lem: interpolation} and \cref{lem: reconstruction}. 
\end{proof}

\subsection{Another view of the algorithm}\label{sec: alternative view of the algorithm}
We now discuss an alternative description of the decoding algorithm. In essence, this is just a rewording of the previous algorithm, but appears to have some qualitative advantages. For instance, the description itself seems simpler here as we don't need to introduce the $\vz$ variables, but instead, end up working with a system of  equations over the original field $\F$ itself.  Moreover, the runtime analysis of the algorithm gives a slightly better bound of $\poly(|S|^k, d^k)$  on the number of field operations needed by the decoding algorithm as opposed to the bound of $\poly(|S|^{k^2}, d^{k^2})$ that is claimed in \cref{thm:main}. 

Given the received word $P: S^k \to \F^{\binom{s + k-1}{k}}$,  we assume that the coordinates of $\F^{\binom{s + k-1}{k}}$ are indexed by $k$-variate monomials of degree at most $s-1$. Let $t = 10m^{k+1}$ and for each $i \in [s]$ and $j \in [t]$, let $\va_{i, j} \in \F^{\binom{i-2 + k}{k}}$ be vectors such that for every $i$, the dimension of the space spanned by $\{\va_{i, 1}, \va_{i,2}, \ldots, \va_{i, t} \}$ over $\F$ equals ${\binom{i-2 + k}{k}}$. Again we think of the coordinates of $\va_{i, j}$ as being indexed by $k$-variate monomials of degree equal to $i-1$. 

Now, from $P$, we construct $P_1, P_2, \ldots, P_t$ where each $P_j$ is a function $S^k$ to $\F^{s}$, such that for every $\vb \in S^k$, the $i^{th}$ coordinate of $P_j(\vb)$ equals the weighted linear combination of the coordinates of $P(\vb)$ indexed by monomials of degree exactly $i-1$, with weights according to $\va_{i, j}$. In other words, the $i^{th}$ coordinate of $P_j(\vb)$ equals 
\[
\sum_{\ve \in \Z_{\geq 0}^k, \|\ve\|_1 = i-1} \va_{i, j}(\ve)\cdot P(\vb)_{\ve} \, ,
\] 
where $P(\vb)_{\ve}$ is the coordinate of $P(\vb)$ indexed by $\ve$.
Now, for the interpolation step, for each $j \in [t]$, we find a polynomial $\tilde{Q}_j = \sum_{i = 1}^m \tilde{Q}_{i, j}(\vx)y_j$ of not too high degree which explains $P_j$ in the sense of \cref{lem: interpolation}. Note that each $\tilde{Q}_j$ is now a polynomial over the original field $\F$. An immediate instantiation of \cref{lem: close enough code words satisfy the equation} for this setting shows that if $f \in \F[\vx]$ of degree at most $d$  and $\enc(f)$ and $P$ are close enough, then for every $j \in [t]$, $\tilde{Q}_j(\vx, \Psi_j(f))$ must be identically zero, where $\Psi_j(f) = \left(\Psi_{j, 1}(f), \ldots, \Psi_{j, m}(f) \right)$ is defined as 
\[
\Psi_{j, i}(f) = \sum_{\ve \in \Z_{\geq}^k, \|\ve\|_1 = i-1} \va_{i, j}(\ve)\cdot \frac{\hpartial  f}{\hpartial \vx^{\ve}} \, .
\]

Before going to the reconstruction step, we note that it might be the case that $\tilde{Q}_1, \tilde{Q}_2, \ldots, \tilde{Q}_t$ depend on different subsets of $\vy$ variables. But since $t > m^{k+1}$, by averaging, it follows that there exist an $\ell \in [m]$  such that at least $m^k$ of the polynomials $\{\tilde{Q}_j : j \in [t]\}$ have the property that they depend on $y_{\ell}$ and do not depend on $y_{\ell'}$ for any $\ell' > \ell$. For the ease of notation, let us assume that $\tilde{Q}_1, \tilde{Q}_2, \ldots, \tilde{Q}_{t'}$ depend on $y_m$, where $t' = m^k$. 

Now, to recover $f$, we solve the equations $\tilde{Q}_j(\vx, \Psi_j(f)) \equiv 0$ for all $j \in [t']$. We solve for $f$ one homogeneous component as in the proof of \cref{lem: reconstruction}. Assuming that we have recovered homogeneous components of degree at most $u$ of $f$, we can follow the proof of \cref{lem: reconstruction} to recover $\Psi_{j, m}(f_{u+1})$ for every $j \in [t']$, where $f_{u+1}$ is the homogeneous component of $f$ of degree $u+1$.\footnote{We might have to do an initial translation of coordinates as in the proof of \cref{lem: reconstruction}.} At this point, the choice of the vectors $\va_{i, j}$, the definition of $\Psi_{j, m}(f_{u+1})$ and the fact that $t' \geq m^k$, we get that we have sufficiently many linearly independent homogeneous linear equations in all the partial derivatives of $f_{u+1}$ of order $(m-1)$. Thus, we can solve this linear system to recover each of these partial derivatives and combine them according to \cref{lem: useful props of delta} to obtain $f_{u+1}$, and proceed to the next step. Moreover, as in \cref{lem: reconstruction}, if we start from the correct coefficients of $f$ in the base case of this process, each of the subsequent steps are unique. 


Thus, instead of working with a single polynomial equation as in a standard application of the polynomial method, this algorithm proceeds via working simultaneously with many equations. 

We now remark on the running time. 
\begin{remark}\label{rmk: improved running time}
We note that in algorithm sketched above, the number of field operations needed is upper bounded by $\poly(|S|^k, d^k)$. This follows from the observation that in this algorithm we are essentially solving $m^k < d^k$ linear systems of size $\poly(|S|^k, d^k)$ over the underlying field $\F$ to recover all codewords close to the received word. 
\end{remark}

\section{Reducing the list size to a constant}\label{sec: constant list size}

In this section, we use the pruning algorithm due to Kopparty, Ron-Zewi, Saraf and Wootters \cite{KoppartyRSW2018} together with \cref{lem: subspace reconstruction} to obtain a shorter list of correct polynomials, thereby improving the bound on the list size in \cref{thm:main} from a polynomial (in the input size) to an absolute constant depending only on the parameter $\epsilon$ and dimension $k$. This would complete the proof of \cref{thm: constant list size}. The first step towards the goal of recovering codewords from a small linear space is the following theorem, which is a natural multivariate analog of \cite[Theorem~17]{GuruswamiK2016} in the work of Guruswami and Kopparty \cite{GuruswamiK2016}. Our proof is essentially the same, apart from the fact that we are in the multivariate setting and hence have to work with Generalized Wronskians matrices. 
\begin{theorem}[subspace restrictions]\label{thm: subspace restrictions}
Let $\F$ be a field of characteristic zero or larger than $d$. Let $\mu \geq w \in \N$ be parameters and let $W \subseteq \F[\vx]$ be an $\F$-linear subspace of $k$-variate polynomials of degree at most $d$, such that dimension of $W$ is at most $w$. For any $\va \in \F^k$, let $H_{\va}$ be the $\F$-linear space of polynomials of degree at most $d$ which vanish with multiplicity at least $\mu$ at $\va$. Then, for every set $S \subseteq \F$, we have, 
\[
\sum_{\va \in S^{k}} \dim(H_{\va} \cap W) \leq \frac{dw |S|^{k-1}}{(\mu - w + 1)}\, .
\] 
\end{theorem}
We use this statement in our proof in this section, and prove it in \cref{sec: wronskian}.

\subsection{The pruning algorithm}
The input to this algorithm is a received word $P$ and a linear subspace $W$ of polynomials of degree at most $d$ in $\F[\vx]$ of dimension at most $w$. The goal is to output all polynomials in $f \in W$ such that $\enc_{s, S}(f)$  agrees with $P$ on at least $\alpha = \delta +\epsilon$ locations. The description of the algorithm has a parameter $r$, which we later set to an appropriate value. 

\paragraph*{Algorithm A}
\begin{enumerate}
\item Choose $\va_1, \va_2, \ldots, \va_{r}$ independently and uniformly at random from $S^k$.  
\item If there is a unique polynomial $f \in  W$ such that $\enc_{s, S}(f)$ and $P$ agree on each of $\va_1, \ldots, \va_{r}$, then output $f$. 
\end{enumerate}
Clearly, the second step of the algorithm can be implemented efficiently via Gaussian Elimination. 

The final pruning algorithm invokes Algorithm A multiple times and outputs the union of all the lists. In the rest of this section, we show that with high probability, this will output the list of  all codewords close to the received word that are contained in the input linear space. 

The algorithm and the analysis is precisely the same as that in the work of Kopparty, Ron-Zewi, Saraf and Wootters~\cite{KoppartyRSW2018}, apart from the fact that we invoke it for multivariate multiplicity codes, whereas in \cite{KoppartyRSW2018} it was designed for folded Reed Solomon Codes and univariate multiplicity codes. We briefly sketch some of the details in the rest of this section. For brevity, we again use $\enc()$ for $\enc_{s, S}()$. We also assume that the dimension $w$ of $W$ is less than the multiplicity parameter $s$ of the code. 

\begin{lemma}[Analogous to {\cite[Lemma~IV.5 (conference version)]{KoppartyRSW2018}}]\label{lem: algorithm A}
For any polynomial $f \in  W$ such that $\dist(\enc(f), P) < \alpha $, $f$ is output by Algorithm A with probability at least 
\[
(1-\alpha)^{r} - w\left(\frac{d}{|S|(s-w)}\right)^r \, .
\]
Moreover, Algorithm A runs in polynomial time in the input size.
\end{lemma}
\begin{proof}[Proof Sketch]
The proof of the lemma is precisely the same as that of \cite[Lemma~IV.5 (conference version)]{KoppartyRSW2018} except we use \cref{thm: subspace restrictions} as opposed to an analogous statement for folded Reed Solomon codes. 
\end{proof}
We are now ready to prove \cref{thm: constant list size}. 


\begin{proof}[Proof of \cref{thm: constant list size}]
Given the error parameter $\epsilon$ and the number of variables $k$, we choose $s, m$ as follows. 
\begin{itemize}
\item $m = \left(\frac{20}{\epsilon}\right)^k$\,,
\item $s = \frac{4}{\epsilon} \cdot \binom{m+k}{k}$\,.
\end{itemize}
We note that for this choice of parameters, $\frac{m}{s-m} < \frac{\epsilon}2$ and hence,
\[
\frac{10}{m^{1/k}} + \frac{m}{s-m} < \epsilon \, ,
\]
as is needed to invoke \cref{lem: interpolation}. We now use \cref{lem: interpolation} to construct the polynomial $Q$ which \emph{explains} the received word $P$, and \cref{lem: subspace reconstruction} to give us a subspace $W$ of polynomials in $\F[\vx]$ of dimension at most  $w = \binom{m+k}{k}$ over $\F$, that  contains all polynomials $f \in \F[\vx]$ of degree at most $d$ such that $\dist(\enc(f), P)<(\delta - \epsilon)$, where $\delta  = 1- d/(s|S|)$ is the relative distance of the code. Let the parameter $r$ be set as
\[
r = \frac{\log (2\cdot\binom{m+k}{k})}{\log (1 + \epsilon/4)} \leq O\left(\frac{k^2\log 1/\epsilon}{\epsilon}\right) \, .
\]
We now instantiate \cref{lem: algorithm A} with inputs being the received word $P$, the subspace $ W$ of dimension at most $w = \binom{m + k}{k}$ and the parameter $r$ as set above. 

A single run of Algorithm A returns at most one polynomial $f$ in $ W$ such that $\dist(\enc(f), P)<(\delta - \epsilon)$. Moreover, every such $f$  is output with probability at least
\[
 \rho = (1 -\delta + \epsilon)^{r} -  w\left(\frac{d}{|S|(s-w)}\right)^r \,.
\]
To simplify this, we note that from the choice of parameters
\begin{align*}
 w\left(\frac{d}{|S|(s-w)}\right)^r &= \binom{m+k}{k}\left(\frac{s}{(s-w)} \cdot (1-\delta)\right)^r \\
 &\leq   \frac{1}{2}\cdot (1+\epsilon/4)^r \left(\frac{1}{(1-\epsilon/4)} \cdot (1-\delta)\right)^r \quad \quad \quad \text{[plugging in the values of $s$, $r$]} \\
  &\leq  \frac{1}{2}\cdot \left(\frac{1+\epsilon/4}{1-\epsilon/4} \cdot (1-\delta)\right)^r \\
  &\leq  \frac{1}{2} \cdot (1-\delta + \epsilon)^r  \, ,
\end{align*}
where the last inequality follows from the fact that $\frac{1+\epsilon/4}{1-\epsilon/4} \cdot (1-\delta) \leq (1-\delta + \epsilon)$, whenever $1 + \delta -\epsilon/2 > 0$, which is always true in our setting, since $\delta, \epsilon \in (0, 1)$. Thus, we get 
\[
\rho \geq \frac12(1-\delta + \epsilon)^r \, .
\] 
Hence, the number of polynomials in the space $W$ such that $\dist(\enc(f), P)<(\delta - \epsilon)$ is at most $\frac1\rho = \frac{2}{(1-\delta + \epsilon)^r}$.

It follows from a union bound that if we run Algorithm A about $O\left(\frac1\rho\cdot \log \frac1\rho\right)$ times with fresh randomness each time, and output every polynomial obtained, with high probability, we would have output all the polynomials $f$ in $ W$ with $\dist(\enc(f), P)<(\delta - \epsilon)$. Thus the number of runs of Algorithm~A is 
\[
O\left(\frac1\rho\cdot \log \frac1\rho\right) = O\left(\frac{r\log(\frac{1}{1-\delta + \epsilon})}{(1-\delta + \epsilon)^r} \right) \leq  \exp\left({O\left(\frac{k^2}{\epsilon}\log^3 \frac{1}{\epsilon}\right)}\right)  \, .
\]
The upper bound on the running time immediately follows from the running time guarantees in \cref{lem: interpolation}, \cref{lem: subspace reconstruction} and the final pruning that happens in the process of recovering the relevant codewords from the subspace output by \cref{lem: subspace reconstruction}. 
\end{proof}

\section{Subspace restrictions of multivariate multiplicity codes}\label{sec: wronskian}
In this section, we prove \cref{thm: subspace restrictions}. For the proof, we  follow the outline of Guruswami and Kopparty~\cite{GuruswamiK2016} and essentially observe that (almost) everything works even for multivariate polynomials. The only difference is that instead of the Wronskian criterion for univariate polynomial, we need to work with the following generalized Wronskian criterion for multivariate polynomials. 
\begin{theorem}[generalized Wronskian criterion]
\label{thm: generalized wronskian}
Let $f_1, f_2, \ldots, f_w \in \F[\vx]$ be $k$-variate polynomials of maximum individual degree at most $d$. If the characteristic of $\F$ is zero or larger than $d$, then the following is true. 
 $f_1, f_2, \ldots, f_w$ are linearly independent over $\F$ if and only if there exist monomials $\vx^{\ve_1}, \vx^{\ve_2}, \ldots, \vx^{\ve_{w}}$ such that for every $i \in [w]$, $\deg(\vx^{\ve_i}) \leq i-1$, and the $w\times w$ matrix $M_{\left(\ve_1, \ldots, \ve_{w}\right)}$ whose $(i,j)$ entry equals $\frac{\hpartial f_j}{\hpartial \vx^{\ve_i}}$ is full rank over the field $\F(\vx)$. 
\end{theorem}

\noindent
The classical Wronskian criterion (and its generalized counterpart) are typically proved for fields of characteristic zero and with the usual notion of partial derivatives (cf., Bostan and Dumas~\cite[Theorem~3]{BostanD2010}). These proofs extend to the above setting. For the sake of completeness, we provide an alternative proof of the above theorem 
in \autoref{sec: proof of gen wronskain}.

Equipped with this criterion, we are now ready to prove \cref{thm: subspace restrictions}
\begin{proof}[Proof of \cref{thm: subspace restrictions}]
Let $f_1, f_2, \ldots, f_w \in W$ be linearly independent polynomials of degree at most $d$ which span $W$. Let $E$ be a subset of $\mu$-tuples of monomials defined as follows. 
\[
E := \{(\vx^{\ve_1}, \vx^{\ve_2}, \ldots, \vx^{\ve_{\mu}}) : \deg(\vx^{\ve_i}) \leq i-1\} \, .
\]
For every $\psi = (\vx^{\ve_1}, \vx^{\ve_2}, \ldots, \vx^{\ve_{\mu}})$ in $E$, let $M_{\psi} \in \F[\vx]^{\mu\times w}$ matrix defined as follows. 
\[
M_{\psi} := 
\begin{bmatrix}
    \frac{\hpartial f_1}{\hpartial \vx^{\ve_1}} & \frac{\hpartial f_2}{\hpartial \vx^{\ve_1}} &  \dots  & \frac{\hpartial f_w}{\hpartial \vx^{\ve_1}} \\
        \frac{\hpartial f_1}{\hpartial \vx^{\ve_2}} & \frac{\hpartial f_2}{\hpartial \vx^{\ve_2}} &  \dots  & \frac{\hpartial f_w}{\hpartial \vx^{\ve_2}} \\
     \vdots & \vdots & \vdots & \vdots  \\
     \vdots & \vdots & \vdots & \vdots  \\
   \frac{\hpartial f_1}{\hpartial \vx^{\ve_{\mu}}} & \frac{\hpartial f_2}{\hpartial \vx^{\ve_{\mu}}} &  \dots  & \frac{\hpartial f_w}{\hpartial \vx^{\ve_{\mu}}}
\end{bmatrix}\, .
\]
And, let $\tilde{M}_{\psi}$ denote the $w\times w$ submatrix of $M_{\psi}$ by taking the first $w$ rows and columns, i.e.,  
\[
\tilde{M}_{\psi} := 
\begin{bmatrix}
    \frac{\hpartial f_1}{\hpartial \vx^{\ve_1}} & \frac{\hpartial f_2}{\hpartial \vx^{\ve_1}} &  \dots  & \frac{\hpartial f_w}{\hpartial \vx^{\ve_1}} \\
        \frac{\hpartial f_1}{\hpartial \vx^{\ve_2}} & \frac{\hpartial f_2}{\hpartial \vx^{\ve_2}} &  \dots  & \frac{\hpartial f_w}{\hpartial \vx^{\ve_2}} \\
     \vdots & \vdots & \vdots & \vdots  \\
     \vdots & \vdots & \vdots & \vdots  \\
   \frac{\hpartial f_1}{\hpartial \vx^{\ve_{w}}} & \frac{\hpartial f_2}{\hpartial \vx^{\ve_{w}}} &  \dots  & \frac{\hpartial f_w}{\hpartial \vx^{\ve_{w}}}
\end{bmatrix}\, .
\]
From \cref{thm: generalized wronskian}, we know that there exists $\psi_0$ in $E$ such that $\tilde{M}_{\psi_0}$ (and hence, $M_{\psi_0}$) is full rank over $\F(\vx)$. Let $L_{\psi_0}$ denote the determinant of  $\tilde{M}_{\psi_0}$. Clearly, $L_{\psi_0}$ is a non-zero $k$-variate polynomial of degree at most $dw$. We note that for many choices of $\psi \in E$, the corresponding matrix $M_{\psi}$ could be of rank less than $w$. Perhaps somewhat surprisingly, all these matrices play a role in the proof. The proof essentially follows from the following claim.

\begin{claim}\label{clm: second claim}
For every $\va \in \F^k$, the multiplicity of $L_{\psi_0}(\vx)$ at $\va$ is at least $(\mu - w + 1)\dim(H_{\va} \cap W)$. 
\end{claim}

We first complete the proof of the theorem using the above claim and then prove the claim.

From \autoref{clm: second claim}, we get 
\[
\sum_{\va \in S^k} (\mu - w + 1)\dim(H_{\va} \cap W) \leq {\sum_{\va \in S^k}} \mult(L(\vx), \va) \, . 
\] 
From the earlier discussion, $L_{\psi_0}$ is a non-zero polynomial of degree at most $dw$. Thus, by \cref{lem: PIL}, the quantity ${\sum_{\va \in S^k}} \mult(L(\vx), \va)$ is upper bounded by $dw|S|^{k-1}$, and this completes the proof of \cref{thm: subspace restrictions}.
\end{proof}

We now prove the claim. For this, we need the following claim.

\begin{claim}\label{clm: first claim}
For every $\psi \in E$, and for every $\va \in \F^k$, 
\[
\rank(M_{\psi}(\va)) \leq w - \dim(H_{\va}\cap W) \, . 
\]
\end{claim}
\begin{proof}[Proof of {\autoref{clm: first claim}}]
We just follow the definition. 
\begin{align*}
\dim(H_{\va} \cap W) &= \dim\left( \set{\vb=(b_1, b_2, \ldots, b_w) \in \F^{w} : \mult(\sum_{i=1}^w b_if_i,\va)\geq \mu}\right) \\
&= \dim \left( \set{\vb=(b_1, b_2, \ldots, b_w) \in \F^{w} : \forall \vx^{\ve} \text{ s.t } \deg(\vx^{\ve}) < \mu,  \sum_{i=1}^w b_i\frac{\hpartial f_i}{\hpartial \vx^{\ve}}(\va) = 0} \right) \\
&=\dim\left(\set{\vb=(b_1, b_2, \ldots, b_w) \in \F^{w} : \forall \psi \in E, (M_{\psi}(\va))\vb = \mathbf{0}} \right) \\
&\leq  \min_{\psi \in E} (\dim(\text{Kernel}(M_{\psi}(\va)))) \\
&\leq  \min_{\psi \in E} (w - \rank(M_{\psi}(\va))) \enspace . \qedhere
\end{align*}
\end{proof}

\begin{proof}[Proof of \autoref{clm: second claim}]
To show the claim, we show that for every monomial $\vx^{\vf}$ of degree less than $(\mu - w + 1)\dim(H_{\va} \cap W)$, the Hasse derivative $\frac{\hpartial L_{\psi_0}}{\hpartial \vx^{\vf}}$ is zero at $\va$. Let $\psi_0 = (\ve_1, \ve_2, \ldots, \ve_w)$. Then, we have (using \cref{prop:Prop_of_Hasse_Derv}: \cref{item:hasse-4,item:hasse-5}).
\[
\frac{\hpartial L_{\psi_0}}{\hpartial \vx^{\vf}}(\va) = \sum_{\vu_1 + \vu_2 + \cdots + \vu_w = \vf} \left(\prod_{j\in[w]}\binom{\ve_j+\vu_j}{\vu_j}\right)\det(\tilde{M}_{(\ve_1 + \vu_1, \ldots, \ve_w + \vu_w)})(\va) \, .
\]
Now, we know that $\sum_j \|\vu_j\|_1 < (\mu - w + 1)\dim(H_{\va} \cap W)$, so there are less than $\dim(H_{\va} \cap W)$ values of $j \in \{1, 2, \ldots, w\}$ such that $\|\vu_j\|_1$ is more than $\mu - w$. Moreover, $\|\vu_j\|_1 \leq \mu - w$ implies that $\|\ve_j\|_1 + \|\vu_j\|_1 \leq \mu-1$. Thus, there is a $\psi \in E$, such that there are more than $w - \dim(H_{\va} \cap W)$ rows of the matrix $\tilde{M}_{(\ve_1 + \vu_1, \ldots, \ve_w + \vu_w)}(\va)$ which are also rows in the matrix $M_{\psi}(\va)$. But, from \autoref{clm: first claim}, we know that for every $\psi \in E$, $M_{\psi}(\va)$ has rank at most $w - \dim(H_{\va} \cap W)$. Thus, each of the matrices $\tilde{M}_{(\ve_1 + \vu_1, \ldots, \ve_w + \vu_w)}(\va)$ in the summand above is rank deficient, and hence has determinant zero. 
\end{proof}

\section*{Acknowledgments}
Madhu, Mrinal, and Prahladh thank Swastik Kopparty for many insightful discussions on multiplicity codes and on the results in \cite{KoppartySY2014, Kopparty2015}.


{\small 
\bibliographystyle{prahladhurl}
\bibliography{mult-bib.bib}
}
\appendix
\section{Exponential number of codewords at a distance $\delta$}\label{sec: super polynomial min weight}
Let $T \subseteq S$ be an arbitrary subset of size $d/s$. For a variable $x_1$, consider the polynomial $f(\vx) = \prod_{b \in T}(x_1 - b)^{s-1}$. At every point $\va \in S^k$ such that $a_1 \in T$, $f(\vx)$ vanishes with multiplicity at least $s$. Moreover, the set $\{\va \in S^k: a_1 \in T \} \subseteq S^k$ is of size exactly $\frac{d}{s}|S|^{k-1}$. Thus, the encoding of every polynomial in the set 
\[
{\cal M} = \left\{\prod_{b \in T}(x_1 - b)^{s-1} : \deg(L(\vx)) = 1, T\subseteq S, |T| = d/s \right\}
\]
 under the $k$-variate multiplicity code, with multiplicity parameter $s$ agrees with the encoding of the polynomial $0$ on at least $d/(qs)$ fraction of points, i.e., the relative distance between them is $(1-\delta)$, where $\delta$ is the distance of the code. Moreover, the set ${\cal M}$ is of size $\binom{q}{d/s} $, which is superpolynomially growing in $d$. 
In this sense, the error tolerance of the result in \cref{thm:main} is the best  one could hope for (up to the $\epsilon > 0$ term) if we are hoping for polynomial list size. 

\section{Generalized Wronskian criterion}
\label{sec: proof of gen wronskain}

In this section, we give a proof of the generalized Wronskian criterion in the multivariate setting that works over fields of finite characteristic, and using the notion of Hasse derivatives. 

We first state and prove a proposition which we will  use to prove \autoref{thm: generalized wronskian}. Given a sequence $f_1,f_2,\ldots,f_w$ of $w$ $k$-variate polynomials of individual degree at most $d$ and a sequence $\ve_1,\ve_2,\ldots,\ve_w$ of $w$ monomials, let $M_{(\ve_1,\ldots,\ve_w)}(f_1,\ldots,f_w)$ be the $w\times w$ matrix whose $(i,j)$-th entry is $\frac{\hpartial f_j}{\hpartial \vx^{\ve_i}}$. Let $W_{(\ve_1,\ldots,\ve_w)}(f_1,\ldots,f_w):=\det{(M_{(\ve_1,\ldots,\ve_w)}(f_1,\ldots,f_w))}$: so, $W_{(\ve_1,\ldots,\ve_w)}(f_1,\ldots,f_w) \in \F[\vx]$.

We say that $\vx^{\ve'}\leq \vx^{\ve}$ if $\ve'\leq \ve$, that is, for all $i\in [k]$: $e_i'\leq e_i$. Let $\lesssim$ be the degree-stratified-lexicographic-total order, which is a extension of the $\leq$ ordering: so, for distinct $\ve$ and $\ve'$, we have $\vx^{\ve'}\lesssim \vx^\ve$ iff $\|\ve'\|_1<\|\ve\|_1$ or $\|\ve'\|_1=\|\ve\|_1$ and $e'_i<e_i$ where $i$ is the first index where $e'_i<e_i$.
Also, for a polynomial $f\in \F[\vx]$, let $\widetilde{f}$ denote its monomial of minimum degree under $\lesssim$ if $f$ is non-zero and 0 otherwise. Thus, for every non-zero polynomial $f$ of the form $\sum_\ve \alpha_\ve \cdot \vx^\ve$ with $\alpha_\ve \in \F$, $\widetilde{f}$ is $\vx^{\ve^*}$ where $\vx^{\ve^*}$ is the least monomial among the set of monomials $\{\vx^\ve : \alpha_\ve \neq 0 \}$. For a monomial, $\ell =\vx^\ve$ we denote $\|\ve\|_1$ by $|\ell|$.

\begin{proposition}
~\\
\label{prop:gen_wron}
\begin{enumerate}
    \item{(linear combinations)} For a fixed $i$, let $f_i= \alpha_i f_i' + \sum_{j\neq i}\alpha_j f_j$ where $\alpha_j\in\F$. Then \[W_{(\ve_1,\ldots,\ve_w)}(f_1,\ldots,f_w)= \alpha_i\cdot W_{(\ve_1,\ldots,\ve_w)}(f_1,\ldots,f_{i-1},f_i',f_{i+1},\ldots,f_w).\]
    \label{itm:1_gen_wron}
    \item{(translation)} Let $\vx+\mathbf{1} = (x_1+1,x_2+1,\ldots,x_k+1)$.
    Then \[(W_{(\ve_1,\ldots,\ve_w)}(f_1(\vx),\ldots,f_w(\vx)))(\vx+\mathbf{1})=(W_{(\ve_1,\ldots,\ve_w)}(f_1(\vx+\mathbf{1}),\ldots,f_w(\vx+\mathbf{1})))(\vx).\]
    \label{itm:2_gen_wron}    
    \item{(minimum monomial)} If $W_{(\ve_1,\ldots,\ve_w)}(\widetilde{f_1},\ldots,\widetilde{f_w}) \neq 0$, then
    \[
    \widetilde{W}_{(\ve_1,\ldots,\ve_w)}(f_1,\ldots,f_w) = W_{(\ve_1,\ldots,\ve_w)}(\widetilde{f_1},\ldots,\widetilde{f_w}).
    \]
    \label{itm:3_gen_wron}
\end{enumerate}
\end{proposition}
\begin{proof}
By linearity of Hasse derivatives we have 
\[\frac{\hpartial f_i'}{\hpartial \vx^{\ve}}= \alpha_i \frac{\hpartial f_i}{\hpartial \vx^{\ve}} + \sum_{j\neq i}\alpha_j\frac{\hpartial f_j}{\hpartial \vx^{\ve}}.\]
Hence, $M_{(\ve_1,\ldots,\ve_w)}(f_1,\ldots,f_w)$ and $M_{(\ve_1,\ldots,\ve_w)}(f_1,\ldots,f_{i-1},f_i',f_{i+1},\ldots,f_w)$ are related by column elementary operations. Thus, their determinants are the same  modulo a multiplicative factor of $\alpha_i$. This proves \autoref{itm:1_gen_wron}. 
The proof of \autoref{itm:2_gen_wron} follows from the fact that for any $f\in \F[\vx]$ we have $(\frac{\hpartial f}{\hpartial \vx^{\ve}})(\vx+\mathbf{1})=(\frac{\hpartial f(\vx+\mathbf{1})}{\hpartial \vx^{\ve}})(\vx)$.
Also, \autoref{itm:3_gen_wron} follows directly by expanding out the determinant.
\end{proof}

Equipped with this proposition, we will now show that if $f_1,\ldots,f_w$ are linearly independent over $\F$, then there exist monomials $\vx^{\ve_1},\ldots,\vx^{\ve_w}$ such that $W_{(\vx^{\ve_1},\ldots,\vx^{\ve_w})}(f_1,\ldots,f_w)\neq 0$ and $\deg{(\vx^{\ve_i})}<i$.

\begin{proof}[Proof of \autoref{thm: generalized wronskian}]
Using \cref{prop:gen_wron}-\cref{itm:1_gen_wron} we can WLOG assume that each $f_i$ has a distinct minimum monomial. We can take an appropriate linear combination of the $f_i$s of the form $f_i \leftarrow f_i + \sum_{j\neq i}\alpha_j f_j$ (this preserves linear independence) to clear out a minimum monomial if it repeats.
Hence, the minimal monomials $\widetilde{f}_i$s are all distinct. Further, by reordering if necessary we can assume that $\widetilde{f}_i$s are in increasing order according to $\lesssim$.
Now, using \cref{prop:gen_wron}-\cref{itm:3_gen_wron}, we are left to show that there are  $\vx^{\ve_1},\ldots,\vx^{\ve_w}$ such that $\deg{(\vx^{\ve_i})}<i$ and $W_{(\ve_1,\ldots,\ve_w)}(\widetilde{f_1},\ldots,\widetilde{f_w}) \neq 0$. To show this first we massage the monomials in the following manner.
\begin{enumerate}
    \item Set $t\leftarrow 0$ and for all $i\in [w]$ let $\ell^0_i \leftarrow \widetilde{f}_i$.
    \item While ($\exists i: |\ell^t_i|\geq i$):
    \begin{enumerate}
        \item For all $i$ let $g^{t+1}_i=\ell^{t}_i(\vx+\mathbf{1})$.
        \item Take appropriate linear combinations of the form $g^{t+1}_i\leftarrow g^{t+1}_i-\sum_{j<i}\alpha_j\cdot g^{t+1}_j$ to ensure that all $\widetilde{g}^{t+1}_i$s are distinct.
        \item For all $i$ set $\ell^{t+1}_i\leftarrow \widetilde{g}^{t+1}_i $. Reorder to ensure that $\ell^{t+1}_i$s are in increasing order wrt $\lesssim$.
        \item $t\leftarrow t+1$.
    \end{enumerate}
\end{enumerate}


We will now show that the while loop terminates in at most $w$ steps and at the end we have $|\ell^t_i|<i$ for all $i\in [w]$. 
Suppose we enter the while loop at a particular value of $t$.
Let $i^*$ be the first index such that $|\ell^t_i|\geq i$. Observe that $g_{i^*}^{t+1}$ will include all monomials $\vx^{\ve'}$ such that $\ve'\leq \ve$ where $\ell_{i^*}^t=\vx^\ve$. This is because the characteristic of $\F$ is larger than the maximum individual degree. Hence, at time $t+1$ we will have $|\ell^{t+1}_j|<j$ for all $j\leq i^*$: for $j<i^*$ step $2(b)$ does not increase the degree of $g^{t+1}_j$ and for $j=i^*$ the minimal monomial $\widetilde{g}_{i^*}^{t+1}$ will be of degree less than $i^*$ as $g_{i^*}^{t+1}$ includes a monomial of degree $i^*-1$ which does not occur in any $g_j^{t+1}$ for $j<i^*$.   
Thus at termination, we have $|\ell_i^t|<i$ for all $i\in [w]$ and further the $\ell_i^t$s are all distinct monomials and in increasing order.

Also, by \cref{prop:gen_wron} we have that if $W_{\ve_1,\ldots,\ve_w}(\ell_1^{t+1},\ldots,\ell_w^{t+1})\neq 0$, then, $W_{\ve_1,\ldots,\ve_w}(\ell_1^{t},\ldots,\ell_w^{t})\neq 0$.
 At termination set $\ell_i=\ell^t_i$.
Hence, we are left to show that there are  $\vx^{\ve_1},\ldots,\vx^{\ve_w}$ such that $\deg{(\vx^{\ve_i})}<i$ and $W_{(\ve_1,\ldots,\ve_w)}(\ell_1,\ldots,\ell_w) \neq 0$. 
Towards this end observe that the matrix $M_{(\ell_1,\ldots,\ell_w)}(\ell_1,\ldots,\ell_w)$ is upper triangular with all the diagonal entries as $1$. For contradiction suppose that $i>j$ and $\frac{\hpartial \ell_i}{\hpartial \ell_j}\neq 0$: then $\ell_j>\ell_i$ which is a contradiction. 
Hence, $W_{(\ell_1,\ldots,\ell_w)}(\ell_1,\ldots,\ell_w)=1$. 
Thus, letting $\vx^{\ve_i}=\ell_i$ for all $i\in [w]$ gives us the requisite monomials $\vx^{\ve_i}$.  

The other direction that if there are monomials $\vx^{\ve_1},\ldots,\vx^{\ve_w}$ such that  $W_{(\vx^{\ve_1},\ldots,\vx^{\ve_w})}(f_1,\ldots,f_w)\neq 0$ then $f_1,\ldots,f_w$ are linearly independent, is simpler. Suppose the $f_i$s are linearly dependent and in particular, $\sum_i \alpha_i f_i$ be a non-trivial linear combination which is zero. Due to linearity of Hasse derivatives we have $(\alpha_1,\ldots,\alpha_w)\in \ker{(M_{(\vx^{\ve_1},\ldots,\vx^{\ve_w})}(f_1,\ldots,f_w))}$.
This completes the proof of \autoref{thm: generalized wronskian}.
\end{proof}
    

\end{document}